\documentclass[12pt]{article}

\usepackage{graphicx}
\usepackage{amsfonts}
\usepackage{amssymb}

\def\gtwid{\mathrel{\raise.3ex\hbox{$>$\kern-.75em\lower1ex\hbox{$\sim$}}}}
\def\ltwid{\mathrel{\raise.3ex\hbox{$<$\kern-.75em\lower1ex\hbox{$\sim$}}}}
\def\square{\kern1pt\vbox{\hrule height 1.2pt\hbox{\vrule width 1.2pt\hskip 3pt
   \vbox{\vskip 6pt}\hskip 3pt\vrule width 0.6pt}\hrule height 0.6pt}\kern1pt}

\begin{document}

\begin{titlepage}

\begin{flushright}
UFIFT-QG-16-08 , CCTP-2016-10 \\
CCQCN-2016-147 , ITCP-IPP 2016/08
\end{flushright}

\vskip 2cm

\begin{center}
{\bf Final Thoughts on the Power Spectra of Scalar Potential Models}
\end{center}

\vskip 1cm

\begin{center}
D. J. Brooker$^{1*}$, N. C. Tsamis$^{2+}$ and R. P. Woodard$^{1\dagger}$
\end{center}

\vskip .5cm

\begin{center}
\it{$^{1}$ Department of Physics, University of Florida,\\
Gainesville, FL 32611, UNITED STATES}
\end{center}

\begin{center}
\it{$^{2}$ Institute of Theoretical Physics \& Computational Physics, \\
Department of Physics, University of Crete, \\
GR-710 03 Heraklion, HELLAS}
\end{center}

\vspace{1cm}

\begin{center}
ABSTRACT
\end{center}
We give final shape to a recent formalism for deriving the functional
forms of the primordial power spectra of single-scalar potential models
and theories which are related to them by conformal transformation. An 
excellent analytic approximation is derived for the nonlocal correction 
factors which are crucial to capture the ``ringing'' that can result from 
features in the potential. We also present the full algorithm for using 
our representation, including the nonlocal factors, to reconstruct the 
inflationary geometry from the power spectra.

\begin{flushleft}
\end{flushleft}

\vskip .5cm

\begin{flushleft}
$^{*}$ e-mail: djbrooker@ufl.edu \\
$^{+}$ e-mail: tsamis@physics.uoc.gr \\
$^{\dagger}$ e-mail: woodard@phys.ufl.edu
\end{flushleft}

\end{titlepage}

\section{Introduction}

The simplest models of primordial inflation are based on general relativity 
(for a spacelike metric $g_{\mu\nu}(x)$) plus a single, minimally coupled 
scalar $\varphi(x)$,
\begin{equation}
\mathcal{L} = \frac{R \sqrt{-g}}{16 \pi G} - \frac12 \partial_{\mu} \varphi
\partial_{\nu} \varphi g^{\mu\nu} \sqrt{-g} - V(\varphi) \sqrt{-g} \; .
\label{Lag}
\end{equation} 
A key prediction is the generation of tensor \cite{Starobinsky:1979ty} and 
scalar \cite{Mukhanov:1981xt} perturbations. These are the first observable
quantum gravitational phenomena ever recognized as such \cite{Woodard:2009ns,
Ashoorioon:2012kh,Krauss:2013pha}. They are also our chief means of 
testing the viability of scalar potential models \cite{Polarski:1995zn,
GarciaBellido:1995fz,Sasaki:1995aw}, and of reconstructing $V(\varphi)$ 
\cite{Mukhanov:1990me,Liddle:1993fq,Lidsey:1995np}.

Reconstruction is simplest in terms of the Hubble representation 
\cite{Liddle:1994dx} using the Hubble parameter $H(t)$ and first slow roll 
parameter $\epsilon(t)$ of the homogeneous, isotropic and spatially flat 
background geometry of inflation,\footnote{The connection to the potential
representation is \cite{Tsamis:1997rk,Saini:1999ba,Nojiri:2005pu,
Woodard:2006nt,Guo:2006ab},
\begin{eqnarray}
\varphi_0(t) = \varphi_0(t_i) \pm \int_{t_i}^{t} \!\! dt' H(t')
\sqrt{ \frac{\epsilon(t')}{4 \pi G}} \; \Longleftrightarrow \; t(\varphi)
\quad , \quad 
V(\varphi) = \frac{[3 \!-\! \epsilon(t)] H^2(t)}{8 \pi G} \Biggl\vert_{
t=t(\varphi)} . \nonumber
\end{eqnarray}}
\begin{equation}
ds^2 = -dt^2 + a^2(t) d\vec{x} \!\cdot\! d\vec{x} \quad \Longrightarrow \quad 
H(t) \equiv \frac{\dot{a}}{a} > 0 \quad , \quad \epsilon(t) \equiv 
-\frac{\dot{H}}{H^2} < 1 \; . \label{geometry}
\end{equation}
Let $t_k$ stand for the time of first horizon crossing, when modes of wave
number $k$ obey $k \equiv H(t_k) a(t_k)$. The tensor and scalar power spectra 
take the form of leading slow roll results at $t=t_k$, multiplied by local 
slow roll corrections also at $t=t_k$, times nonlocal factors involving times 
near $t=t_k$ \cite{Brooker:2015iya,Brooker:2016xkx},
\begin{eqnarray}
\Delta^2_{h}(k) & = & \frac{16}{\pi} \, G H^2(t_k) \!\times\! C\Bigl( 
\epsilon(t_k)\Bigr) \!\times\! \exp\Bigl[\tau[\epsilon](k)\Bigr] \; , 
\label{fullDh} \\
\Delta^2_{\mathcal{R}}(k) & = & \frac{G H^2(t_k)}{\pi \epsilon(t_k)} 
\!\times\! C\Bigl( \epsilon(t_k)\Bigr) \!\times\! \exp\Bigl[\sigma[\epsilon](k)
\Bigr] \; . \label{fullDR}
\end{eqnarray}
The local slow roll correction $C(\epsilon)$ is,
\begin{equation}
C(\epsilon) \equiv \frac1{\pi} \Gamma^2\Bigl( \frac12 \!+\! 
\frac1{1 \!-\! \epsilon}\Bigr) \Bigl[ 2 (1 \!-\! \epsilon)
\Bigr]^{\frac2{1-\epsilon}} \approx 1 - \epsilon \; . \label{Cdef}
\end{equation}
The nonlocal correction exponents, $\tau[\epsilon](k)$ and $\sigma[\epsilon](k)$,
vanish for $\dot{\epsilon} = 0$ and effectively depend on the geometry only a few
e-foldings before and after $t_k$ \cite{Brooker:2015iya,Brooker:2016xkx}.

The purpose of this paper is to rationalize and simplify our formalism for
evolving the norms of the mode functions, rather than the mode functions
\cite{Romania:2012tb}, and then to derive an excellent analytic approximation 
for the nonlocal correction exponents $\tau[\epsilon](k)$ and $\sigma[\epsilon](k)$. 
We also demonstrate how this approximation can be used to reconstruct the 
inflationary geometry from the power spectra, even for models which possess 
features. These topics represent sections 2-3, 4 and 5, respectively. In section 
6 we discuss some of the many applications \cite{Brooker:2016imi,Brooker:2016oqa}
this formalism facilitates.

\begin{figure}[ht]
\includegraphics[width=5.0cm,height=4.0cm]{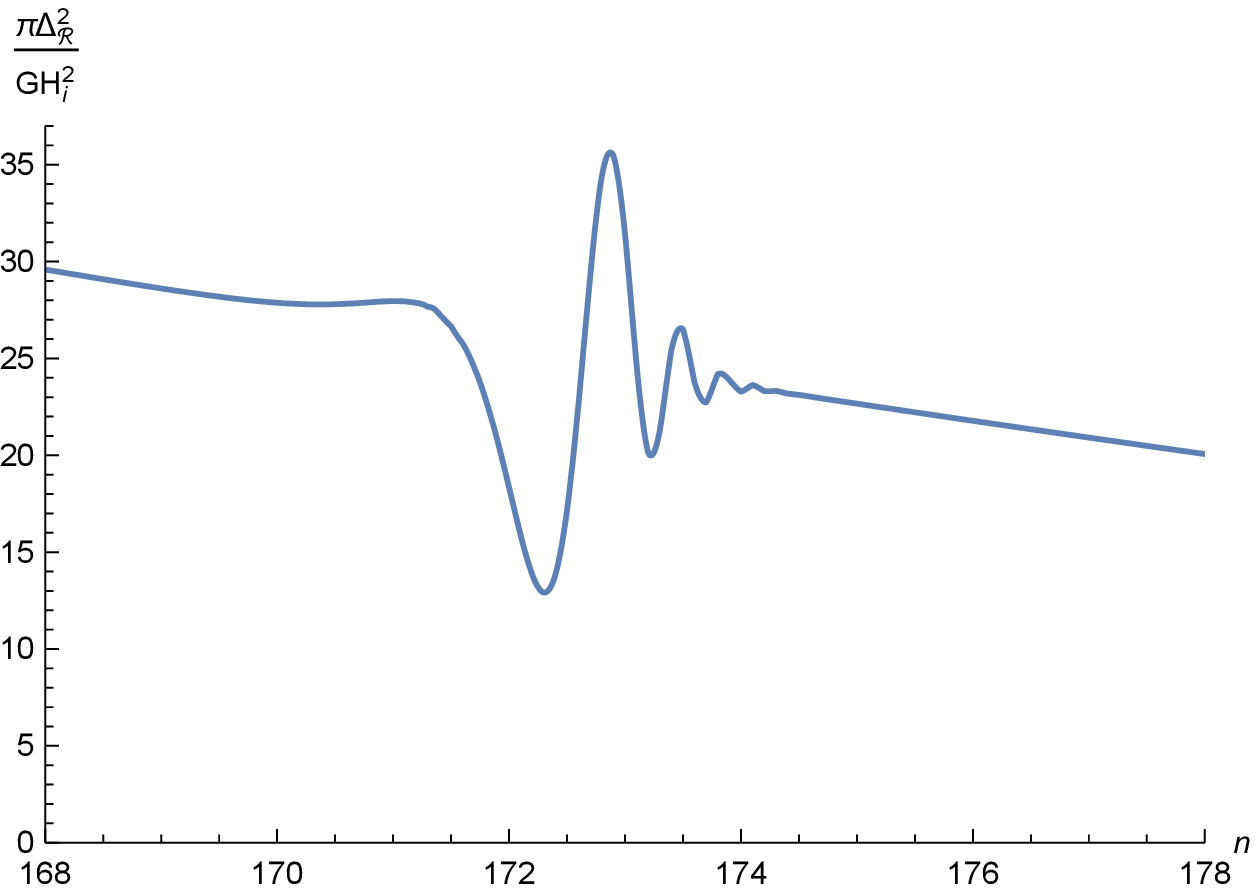}\hskip 2cm
\includegraphics[width=5.0cm,height=4.0cm]{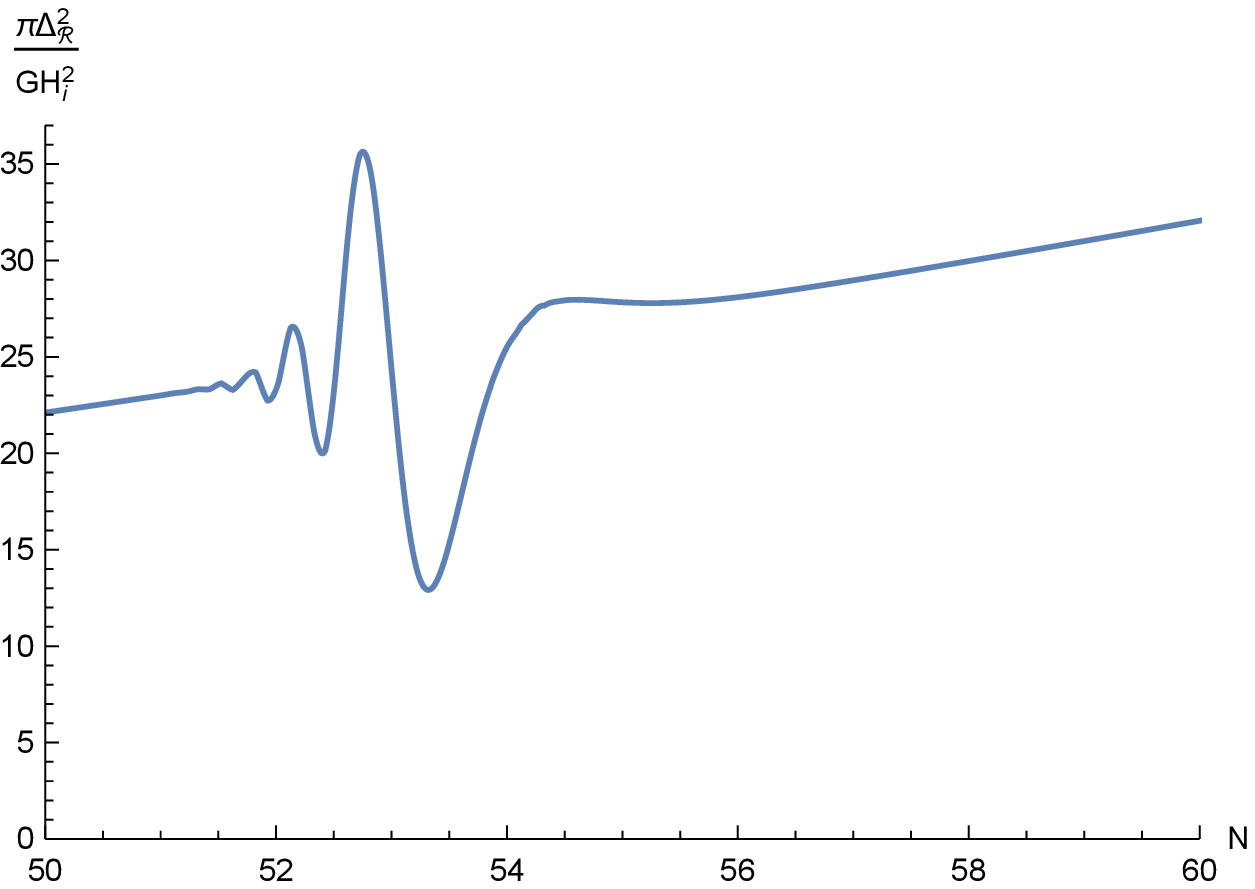}
\caption{The left hand graph shows one model's scalar power spectrum as a 
function of $n$, the number of e-foldings from the beginning of inflation 
to first horizon crossing. The right hand graph shows the same power spectrum 
versus $N$, the number of e-foldings {\it until} the end of inflation. Early 
times correspond to small $n$ and large $N$, whereas late times correspond to
large $n$ and small $N$.}
\label{nversusN}
\end{figure}

We shall often employ the alternate time parameter provided by $n \equiv
\ln[a(t)/a_i]$, the number of e-foldings since inflation's onset. This is 
superior to the co-moving time $t$ by virtue of being dimensionless and 
relating evolution to the size of the universe. We shall abuse the notation 
slightly by writing $H(n)$ and $\epsilon(n)$, instead of the correct but
cumbersome expressions $H(t(n))$ and $\epsilon(t(n))$. Which time parameter 
pertains should be clear from context, and from our exclusive use of $\ell$, 
$m$ and $n$ to stand for e-foldings. Over-dots represent time derivatives 
and primes stand for $n$ derivatives,
\begin{equation}
\epsilon = -\frac{\dot{H}}{H^2} = -\frac{H'}{H} \;\; \Longleftrightarrow \;\; 
H = \frac{H_i}{1 \!+\! \int_{t_i}^{t} \!\! dt' \, \epsilon(t')} =
H_i \exp\Bigl[ -\!\!\int_{0}^{n} \!\!\! dm \, \epsilon(m)\Bigr] \, .
\end{equation}
We caution readers against confusing $n$ with the common parameter $N \equiv 
n_e - n$, the number of e-foldings until the end of inflation (at $n = n_e$). 
Figure~\ref{nversusN} illustrates the difference.

\section{Our Formalism in General}

The tree order tensor power spectrum is obtained by evolving the graviton 
mode function $u(t,k)$ past the time of first horizon crossing
\cite{Mukhanov:1990me,Liddle:1993fq,Lidsey:1995np},
\begin{equation}
\Delta^2_{h}(k) = \frac{k^3}{2 \pi^2} \!\times\! 32\pi G \!\times\! 2 \!\times\!
\lim_{t \gg t_k} \Bigl\vert u(t,k) \Bigr\vert^2 \; . \label{Dhdef}
\end{equation}
We do not possess exact solutions for $u(t,k)$ for realistic geometries
$\epsilon(t)$, but we do know the evolution equation, the Wronskian and the
form at asymptotically early times \cite{Mukhanov:1990me,Liddle:1993fq,
Lidsey:1995np,Woodard:2014jba},
\begin{equation}
\ddot{u} + 3 H \dot{u} + \frac{k^2}{a^2} u = 0 \;\; , \;\; u \dot{u}^* 
\!-\! \dot{u} u^* = \frac{i}{a^3} \;\; , \;\; u(t,k) \longrightarrow 
\frac{ \exp[-i k \!\int_{t_i}^{t} \!\! \frac{dt'}{a(t')} ]}{\sqrt{2 k a^2(t)}}
\; . \label{usualumode}
\end{equation}
Because the power spectrum depends upon the norm-squared, rather than the
rapidly-varying phase, it is better to convert (\ref{usualumode}) into a
nonlinear evolution equation for $M(t,k) \equiv \vert u(t,k)\vert^2$
\cite{Romania:2012tb},
\begin{equation}
\ddot{M} + 3 H \dot{M} + \frac{2 k^2}{a^2} M = \frac1{2M} \Bigl( \dot{M}^2
+ \frac1{a^6}\Bigr) \quad , \quad M(t,k) \longrightarrow \frac1{2 k a^2(t)}
\; . \label{Mtform}
\end{equation}
If necessary, the mode function can be easily recovered \cite{Brooker:2016xkx},
\begin{equation}
u(t,k) = \sqrt{ M(t,k)} \, \exp\Bigl[ -\frac{i}{2} \! \int_{t_i}^{t} \! 
\frac{dt'}{a^3(t') M(t',k)} \Bigr] \; . \label{ufromM}
\end{equation}

Relation (\ref{Mtform}) can be improved by changing to the dimensionless
time parameter $n = \ln[a(t)/a_i]$,
\begin{equation}
\Bigl( \frac{M'}{M}\Bigr)' + \frac12 \Bigl( \frac{M'}{M}\Bigr)^2 + (3 \!-\!
\epsilon) \frac{M'}{M} + \frac{2 k^2}{H^2 a^2} - \frac1{2 H^2 a^6 M^2} = 0 \; .
\label{Mnform}
\end{equation}
A further improvement comes by factoring out an (at this stage) arbitrary 
approximate solution, $M_0(t,k)$, to derive a damped, driven oscillator
equation (with small nonlinearities) for the residual exponent
\cite{Brooker:2015iya},
\begin{equation}
M = M_0 \!\times\! e^{-\frac12 h} \quad \Longrightarrow \quad h'' - 
\frac{\omega'}{\omega} h' + \omega^2 h = S_h + \frac14 {h'}^2 - \omega^2 
\Bigl[e^h \!-\! 1 \!-\! h\Bigr] \; . \label{heqn}
\end{equation}
Here the frequency $\omega(n,k)$ and the tensor source $S_h(n,k)$ are,
\begin{equation}
\omega \equiv \frac1{H a^3 M_0} \quad \Longrightarrow \quad S_h = -2 
\Bigl(\frac{\omega'}{\omega}\Bigr)' + \Bigl( \frac{\omega'}{\omega}\Bigr)^2
+ 2 \epsilon' - (3 - \epsilon)^2 + \frac{4 k^2}{H^2 a^2} - \omega^2 
\; . \label{tensorsource}
\end{equation}
It is an amazing fact that an exact Green's function exists for the 
left hand side of equation (\ref{heqn}), valid for {\it any} choice of the
approximate solution $M_0$ \cite{Brooker:2015iya},
\begin{equation}
G_h(n;m) = \frac{\theta(n \!-\! m)}{\omega(m,k)} \, \sin\Biggl[ 
\int_{0}^{n} \!\!\! d\ell \, \omega(\ell,k) \Biggr] \; . 
\label{tensorgreen}
\end{equation}
This permits us to solve (\ref{heqn}) perturbatively $h = h_1 + h_2 + \dots$
by expanding in the nonlinear terms,
\begin{eqnarray}
h_1(n,k) & = & \int_0^{n} \!\!\! dm \, G_h(n;m) S_h(m,k) \; , \\
h_2(n,k) & = & \int_0^{n} \!\!\! dm \, G_h(n;m) \Biggl\{ \frac14 \Bigl[h_1'(m,k)
\Bigr]^2 - \frac12 \Bigl[\omega(m,k) h_1(m,k)\Bigr]^2 \Biggr\} \; . \qquad 
\end{eqnarray}

The tree order scalar power spectrum is obtained by evolving the $\zeta$ 
mode function $v(t,k)$ past the time of first horizon crossing
\cite{Mukhanov:1990me,Liddle:1993fq,Lidsey:1995np},
\begin{equation}
\Delta^2_{\mathcal{R}}(k) = \frac{k^3}{2 \pi^2} \!\times\! 4\pi G \!\times\!
\lim_{t \gg t_k} \Bigl\vert v(t,k) \Bigr\vert^2 \; . \label{DRdef}
\end{equation}
Just as for its tensor cousin, we lack exact solutions for $v(t,k)$ for realistic 
geometries $\epsilon(t)$, but we do know the evolution equation, the Wronskian 
and the form at asymptotically early times \cite{Mukhanov:1990me,Liddle:1993fq,
Lidsey:1995np,Woodard:2014jba},
\begin{equation}
\ddot{v} + \Bigl(3 H \!+\! \frac{\dot{\epsilon}}{\epsilon} \Bigr) \dot{v} + 
\frac{k^2}{a^2} v = 0 \;\; , \;\; v \dot{v}^* \!-\! \dot{v} v^* = 
\frac{i}{\epsilon a^3} \;\; , \;\; v(t,k) \longrightarrow \frac{ \exp[-i k 
\!\int_{t_i}^{t} \!\! \frac{dt'}{a(t')} ]}{\sqrt{2 k \epsilon(t) a^2(t)}} \; . 
\label{usualvmode}
\end{equation}
Converting to the norm-squared $\mathcal{N}(t,k) \equiv \vert v(t,k)\vert^2$ 
gives \cite{Brooker:2016xkx},
\begin{equation}
\ddot{\mathcal{N}} + \Bigl(3 H \!+\! \frac{\dot{\epsilon}}{\epsilon}\Bigr) 
\dot{\mathcal{N}} + \frac{2 k^2}{a^2} \mathcal{N} = \frac1{2\mathcal{N}} 
\Bigl( \dot{\mathcal{N}}^2 + \frac1{\epsilon^2 a^6}\Bigr) \quad , \quad 
\mathcal{N}(t,k) \longrightarrow \frac1{2 k \epsilon(t) a^2(t)} \; . 
\label{Ntform}
\end{equation}
The scalar mode function mode can be recovered from $\mathcal{N}(t,k)$
\cite{Brooker:2016xkx},
\begin{equation}
v(t,k) = \sqrt{\mathcal{N}(t,k)} \, \exp\Bigl[ -\frac{i}{2} \! \int_{t_i}^{t} \! 
\frac{dt'}{\epsilon(t') a^3(t') \mathcal{N}(t',k)} \Bigr] \; . \label{vfromN}
\end{equation}

Converting from co-moving time $t$ to $n = \ln[a(t)/a_i]$ gives,
\begin{equation}
\Bigl( \frac{\mathcal{N}'}{\mathcal{N}}\Bigr)' + \frac12 \Bigl( 
\frac{\mathcal{N}'}{\mathcal{N}}\Bigr)^2 + \Bigl(3 \!-\! \epsilon \!+\!
\frac{\epsilon'}{\epsilon}\Bigr) \frac{\mathcal{N}'}{\mathcal{N}} + 
\frac{2 k^2}{H^2 a^2} - \frac1{2 \epsilon^2 H^2 a^6 \mathcal{N}^2} = 0 \; . 
\label{Nnform}
\end{equation}
Factoring out by an arbitrary approximate solution $\mathcal{N}_0(t,k)$ 
produces another damped, driven oscillator equation for the residual exponent,
\begin{equation}
\mathcal{N} = \mathcal{N}_0 \!\times\! e^{-\frac12 g} \quad \Longrightarrow 
\quad g'' - \frac{\Omega'}{\Omega} g' + \Omega^2 g = S_g + \frac14 {g'}^2 - 
\Omega^2 \Bigl[e^g \!-\! 1 \!-\! g\Bigr] \; . \label{geqn}
\end{equation}
Here the frequency $\Omega(n,k)$ and the scalar source $S_g(n,k)$ are,
\begin{eqnarray}
\Omega & \equiv & \frac1{\epsilon H a^3 \mathcal{N}_0} \; , \\
S_g & = & -2 \Bigl(\frac{\Omega'}{\Omega}\Bigr)' + \Bigl( \frac{\Omega'}{\Omega}
\Bigr)^2 + 2 \epsilon' - \Bigl(3 \!-\! \epsilon \!+\! \frac{\epsilon'}{\epsilon}
\Bigr)^2 - 2 \Bigl( \frac{\epsilon'}{\epsilon}\Bigr)' + \frac{4 k^2}{H^2 a^2} 
- \Omega^2 \; . \qquad \label{scalarsource}
\end{eqnarray}
Making the replacement $\omega \rightarrow \Omega$ in (\ref{tensorgreen}) gives
an exact Green's function which is valid for any choice of $\mathcal{N}_0$,
\begin{equation}
G_g(n;m) = \frac{\theta(n \!-\! m)}{\Omega(m,k)} \, \sin\Biggl[ 
\int_{0}^{n} \!\!\! d\ell \, \Omega(\ell,k) \Biggr] \; . 
\label{scalargreen}
\end{equation}
And we can of course develop a perturbative solution to (\ref{geqn})
$g = g_1 + g_2 + \dots$,
\begin{eqnarray}
g_1(n,k) & = & \int_0^{n} \!\!\! dm \, G_g(n;m) S_g(m,k) \; , \\
g_2(n,k) & = & \int_0^{n} \!\!\! dm \, G_g(n;m) \Biggl\{ \frac14 \Bigl[g_1'(m,k)
\Bigr]^2 - \frac12 \Bigl[\Omega(m,k) g_1(m,k)\Bigr]^2 \Biggr\} \; . \qquad 
\end{eqnarray}

\section{Choosing $M_0(t,k)$ and $\mathcal{N}_0(t,k)$ Effectively}

The formalism of the previous section is valid for all choices of the approximate
solutions $M_0(t,k)$ and $\mathcal{N}_0(t,k)$. Of course the correction exponents
$h(n,k)$ and $g(n,k)$ will be smaller if the zeroth order solutions are more
carefully chosen. In previous work we used the instantaneously constant 
$\epsilon$ solutions \cite{Brooker:2015iya,Brooker:2016xkx},
\begin{equation}
M_{\rm inst}(t,k) \equiv \frac{z(t,k) \mathcal{H}\Bigl(\nu(t),z(t,k)\Bigr)}{
2 k a^2(t)} \; , \; \mathcal{N}_{\rm inst}(t,k) \equiv \frac{z(t,k)
\mathcal{H}\Bigl(\nu(t),z(t,k)\Bigr)}{2 k \epsilon(t) a^2(t)} \; , \label{inst}
\end{equation}
where we define,
\begin{equation}
\mathcal{H}(\nu,z) \equiv \frac{\pi}{2} \Bigl\vert H^{(1)}_{\nu}(z)\Bigr\vert^2
\;\; , \;\; \nu(t) \equiv \frac12 + \frac1{1 \!-\! \epsilon(t)}  
\;\; , \;\; z(t,k) \equiv \frac{k}{[1 \!-\! \epsilon(t)] H(t) a(t)} \; .
\label{Hnuz}
\end{equation}
However, the choice (\ref{inst}) has the undesirable effect of complicating the 
late time limits. The physical quantities $M(t,k)$ and $\mathcal{N}(t,k)$ freeze
in to constant values soon after first horizon crossing, but continued evolution 
in $\epsilon(t)$ prevents $M_0(t,k)$ and $\mathcal{N}_0(t,k)$ from approaching
constants. Hence the residual exponents $h(n,k)$ and $g(n,k)$ must evolve so as
to cancel this effect.

We can make the late time limits simpler by adopting a piecewise choice for the 
approximate solutions,
\begin{eqnarray}
M_0(t,k) & = & \theta(t_k \!-\! t) M_{\rm inst}(t,k) + \theta(t \!-\! t_k) 
\overline{M}_{\rm inst}(t,k) \; , \label{M0actual} \\
\mathcal{N}_0(t,k) & = & \theta(t_k \!-\! t) \mathcal{N}_{\rm inst}(t,k) +
\theta(t \!-\! t_k) \overline{\mathcal{N}}_{\rm inst}(t,k) \; . \label{N0actual} 
\end{eqnarray}
By $\overline{M}_{\rm inst}(t,k)$ and $\overline{N}_{\rm inst}(t,k)$ we mean the
solutions which would pertain for the ersatz geometry,
\begin{equation}
\overline{a}(n) = a(n) = a_k e^{\Delta n} \quad , \quad \overline{H}(n) = H_k
e^{-\epsilon_k \Delta n} \quad , \quad \overline{\epsilon}(n) = \epsilon_k 
\; . \label{fakegeom}
\end{equation}
Here and henceforth $\Delta n \equiv n - n_k$ stands for the number of e-foldings 
from horizon crossing. To be explicit about the over-lined quantities,
\begin{equation}
\overline{M}_{\rm inst} \equiv \frac{\overline{z} \, \mathcal{H}(\nu_k,\overline{z})}{
2 k \overline{a}^2} \quad , \quad \overline{\mathcal{N}}_{\rm inst} \equiv
\frac{\overline{z} \, \mathcal{H}(\nu_k,\overline{z})}{2 k \epsilon_k \overline{a}^2} 
\quad , \quad \overline{z} \equiv \frac{e^{(1-\epsilon_k) \Delta n}}{1 \!-\! 
\epsilon_k} \; .
\end{equation}
With the choice (\ref{M0actual}-\ref{N0actual}) the approximate solutions rapidly
freeze in to constants,
\begin{equation}
\lim_{t \gg t_k} M_0(t,k) = \frac{H_k^2}{2 k^3} \!\times\! C(\epsilon_k) \quad , 
\quad \lim_{t \gg t_k} \mathcal{N}_0(t,k) = \frac{H_k^2}{2 \epsilon_k k^3} \!\times\! 
C(\epsilon_k) \; . \label{M0N0late}
\end{equation}
This establishes the forms (\ref{fullDh}-\ref{fullDR}) for the power spectra and 
fixes the nonlocal correction exponents to,
\begin{equation}
\tau[\epsilon](k) = -\frac12 \lim_{t \gg t_k} g(t,k) \quad , \quad 
\sigma[\epsilon](k) = -\frac12 \lim_{t \gg t_k} h(t,k) \; . \label{fullsigmatau}
\end{equation}

It remains to specialize the sources to (\ref{M0actual}-\ref{N0actual}). First
note the simple relation between the scalar and tensor frequencies, 
\begin{equation}
\Omega(n,k) = \theta(n_n \!-\! n) \omega(n,k) + \theta(n \!-\! n_k) \omega(n,k) 
\!\times\! \frac{\epsilon_k}{\epsilon(n)} \; . \label{freqsactual}
\end{equation}
This means the scalar source (\ref{scalarsource}) consists of the tensor source
(\ref{tensorsource}) minus a handful of terms mostly involving $\epsilon(n)$,
\begin{eqnarray}
\lefteqn{S_g(n,k) = S_h(n,k) - 2\theta(n_k \!-\! n) \Bigl[ \Bigl( 
\frac{\epsilon'}{\epsilon}\Bigr)' + \frac12 \Bigl(\frac{\epsilon'}{\epsilon}
\Bigr)^2 + (3 \!-\! \epsilon)\frac{\epsilon'}{\epsilon}\Bigr] } \nonumber \\
& & \hspace{1.9cm} + 2 \delta(n \!-\! n_k) \frac{\epsilon'}{\epsilon} - 2 
\theta(n \!-\! n_k) \Bigl[\Bigl(3 \!-\! \epsilon \!+\! \frac{\omega'}{\omega}\Bigr) 
\frac{\epsilon'}{\epsilon} + \omega^2 \Bigl(\frac{\epsilon_k^2}{\epsilon^2} 
\!-\! 1\Bigr)\Bigr] \; . \qquad \label{SgfromSh}
\end{eqnarray}

To obtain an explicit formula for the tensor source we first note that the
tensor frequency is,
\begin{equation}
\omega(n,k) = \theta(n_k \!-\! n) \frac{2 (1 \!-\! \epsilon)}{\mathcal{H}(\nu,z)}
+ \theta(n \!-\! n_k) \frac{2 (1 \!-\! \epsilon_k)}{\mathcal{H}(\nu_k,
\overline{z})} \!\times\! \frac{\overline{H}}{H} \; . \label{omegaactual}
\end{equation}
Hence the $n$ derivative of its logarithm is,
\begin{equation}
\frac{\omega'}{\omega} = \theta(n_k \!-\! n) \Bigl[-\frac{\epsilon'}{1 \!-\!
\epsilon} - \frac{\mathcal{H}'}{\mathcal{H}}\Bigr] + \theta(n \!-\! n_k) \Bigl[ 
\Delta \epsilon - \frac{\overline{\mathcal{H}}'}{\overline{\mathcal{H}}} \Bigr] 
\; , \label{omegap}
\end{equation}
where $\Delta \epsilon \equiv \epsilon(n) - \epsilon_k$ and $\overline{
\mathcal{H}} \equiv \mathcal{H}(\nu_k,\overline{z})$. Before horizon crossing
$\nu = \frac12 + \frac1{1-\epsilon}$ is time dependent and $z = k/[(1-\epsilon) 
H a]$ so we have,
\begin{eqnarray}
\lefteqn{\nu' = \frac{\epsilon'}{(1 \!-\! \epsilon)^2} \; , \; z' = 
-\Bigl[1 \!-\! \epsilon \!-\! \frac{\epsilon'}{1 \!-\! \epsilon}\Bigr] z }
\nonumber \\
& & \hspace{4cm} \Longrightarrow -\frac{\mathcal{H}'}{\mathcal{H}} = 
-\frac{\epsilon'}{(1 \!-\! \epsilon)^2} \mathcal{A} + \Bigl[1 \!-\! \epsilon 
\!-\! \frac{\epsilon'}{1 \!-\! \epsilon}\Bigr] \mathcal{B} \; , \qquad  
\label{Hbefore}
\end{eqnarray}
where $\mathcal{A}$ and $\mathcal{B}$ involve derivatives of $\mathcal{H}(\nu,z)$
with respect to $\nu$ and $\zeta = \ln(z)$,
\begin{equation}
\mathcal{A} \equiv \partial_{\nu} \ln\Bigl[\mathcal{H}(\nu,e^{\zeta})\Bigr] 
\quad , \quad \mathcal{B} \equiv \partial_{\zeta} \ln\Bigl[\mathcal{H}(\nu,
e^{\zeta})\Bigr] \; . \label{dH}
\end{equation}
The analogous result after horizon crossing is much simpler,
\begin{equation}
-\frac{\overline{\mathcal{H}}'}{\overline{\mathcal{H}}} = (1 \!-\! \epsilon_k)
\overline{\mathcal{B}} \; , \label{Hafter}
\end{equation}
where $\overline{\mathcal{B}}$ means $\mathcal{B}$ with $\nu$ specialized to 
$\nu_k$ and $e^{\zeta}$ specialized to $\overline{z}$.

Taking the derivative of $\omega'/\omega$ before horizon crossing,
\begin{eqnarray}
\lefteqn{ \Bigl( \frac{\omega'}{\omega}\Bigr)' = -\frac{\epsilon''}{1 \!-\!
\epsilon} \!-\! \frac{{\epsilon'}^2}{(1 \!-\! \epsilon)^2} \!-\! \Bigl[ 
\frac{\epsilon''}{(1 \!-\! \epsilon)^2} \!+\! \frac{2 {\epsilon'}^2}{(1 \!-\! 
\epsilon)^3}\Bigr] \mathcal{A} \!-\! \Bigl[\epsilon' + \frac{\epsilon''}{1 \!-\! 
\epsilon} \!+\! \frac{{\epsilon'}^2}{(1 \!-\! \epsilon)^2} \Bigr] \mathcal{B} } 
\nonumber \\
& & \hspace{2cm} - \frac{{\epsilon'}^2}{(1 \!-\! \epsilon)^4} \, \mathcal{C} +
\frac{2 \epsilon'}{(1 \!-\! \epsilon)^2} \Bigl[1 \!-\! \epsilon \!-\! 
\frac{\epsilon'}{1 \!-\! \epsilon}\Bigr] \mathcal{D} - \Bigl[1 \!-\! \epsilon
\!-\! \frac{\epsilon'}{1 \!-\! \epsilon}\Bigr] \mathcal{E} \; , \qquad 
\label{omegapp}
\end{eqnarray}
requires three second derivatives of $\ln[\mathcal{H}]$, 
\begin{equation}
\mathcal{C} \equiv \partial^2_{\nu} \ln\Bigl[\mathcal{H}(\nu,e^{\zeta})\Bigr] 
\;\; , \;\; \mathcal{D} \equiv \partial_{\zeta} \partial_{\nu} 
\ln\Bigl[\mathcal{H}(\nu,e^{\zeta})\Bigr] \;\; , \;\; \mathcal{E} \equiv 
\partial_{\zeta}^2 \ln\Bigl[\mathcal{H}(\nu,e^{\zeta})\Bigr] \; . \label{ddH}
\end{equation}
Bessel's equation and the Wronskian of $H^{(1)}_{\nu}(e^{\zeta})$ imply,
\begin{equation}
2 (1 \!-\! \epsilon)^2 \mathcal{E} + (1 \!-\! \epsilon)^2 \mathcal{B}^2 
- (3 \!-\! \epsilon)^2 + 4 (1\!-\! \epsilon)^2 e^{2 \zeta} - \Bigl( 
\frac{2 (1 \!-\! \epsilon)}{\mathcal{H}}\Bigr)^2 = 0 \; . \label{Bessel}
\end{equation}
Substituting relations (\ref{omegap}), (\ref{Hbefore}), (\ref{omegapp}) and
(\ref{Bessel}) in the definition of the tensor source (\ref{tensorsource})
gives,
\begin{eqnarray}
\lefteqn{ t < t_k \quad \Longrightarrow \quad S_{\rm before} = 
\frac{2 \epsilon''}{1 \!-\! \epsilon} \Bigl[ 1 \!+\! \frac{\mathcal{A}}{1 
\!-\! \epsilon} \!+\! \mathcal{B} \Bigr] + 2 \epsilon' \Bigl[1 \!-\! 
\frac{\mathcal{A} \mathcal{B}}{1 \!-\! \epsilon} \!-\! \mathcal{B}^2 \!-\! 
\frac{2 \mathcal{D}}{1 \!-\! \epsilon} - 2 \mathcal{E}\Bigr] } \nonumber \\
& & \hspace{0.5cm} + \frac{2 {\epsilon'}^2}{(1 \!-\! \epsilon)^2} \Biggl\{ -
\frac12 + \frac12 \Bigl[2 \!+\! \frac{\mathcal{A}}{1 \!-\! \epsilon} \!+\!
\mathcal{B}\Bigr]^2 \!+\! \frac{\mathcal{A}}{1 \!-\! \epsilon} \!+\! 
\frac{\mathcal{C}}{(1 \!-\! \epsilon)^2} \!+\! \frac{2 \mathcal{D}}{1 \!-\!
\epsilon} \!+\! \mathcal{E}\Biggr\} 
\; . \qquad \label{Shbefore}
\end{eqnarray}
The analogous result after horizon crossing is,
\begin{eqnarray}
\lefteqn{t > t_k \quad \Longrightarrow \quad S_{\rm after} = 2 \Delta \epsilon 
\Bigl[3 \!-\! \epsilon_k \!+\! (1 \!-\! \epsilon_k) \overline{\mathcal{B}}\Bigr] }
\nonumber \\
& & \hspace{6cm} + 4 \Bigl[ \frac{k^2}{\overline{H}^2 a^2} - \Bigl( \frac{1 \!-\! 
\epsilon_k}{\overline{\mathcal{H}}}\Bigr)^2 \Bigr] \Bigl[ \frac{\overline{H}^2}{
H^2} \!-\! 1\Bigr] \; . \qquad \label{Shafter}
\end{eqnarray}
There is also a jump at horizon crossing so that the complete result is,
\begin{equation}
S_h = \theta(n_k \!-\! n) \, S_{\rm before} - \delta(n \!-\! n_k) 
\frac{2 \epsilon'}{1 \!-\! \epsilon} \Bigl[1 \!+\! \frac{\mathcal{A}}{1 \!-\! 
\epsilon} \!+\! \mathcal{B}\Bigr] + \theta(n \!-\! n_k) \, S_{\rm after} \; .
\label{Shcomplete}
\end{equation}  

\begin{figure}[ht]
\includegraphics[width=6.0cm,height=4.8cm]{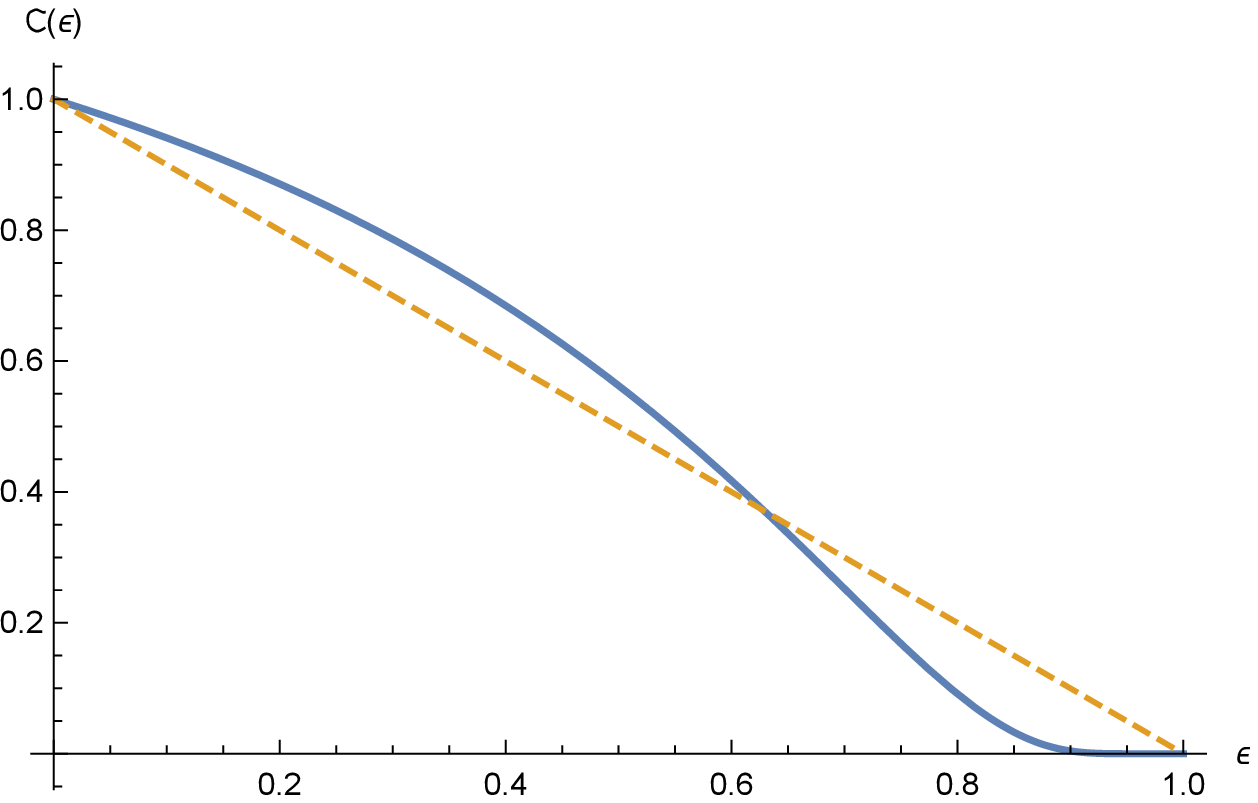}\hskip 1cm
\includegraphics[width=6.0cm,height=4.8cm]{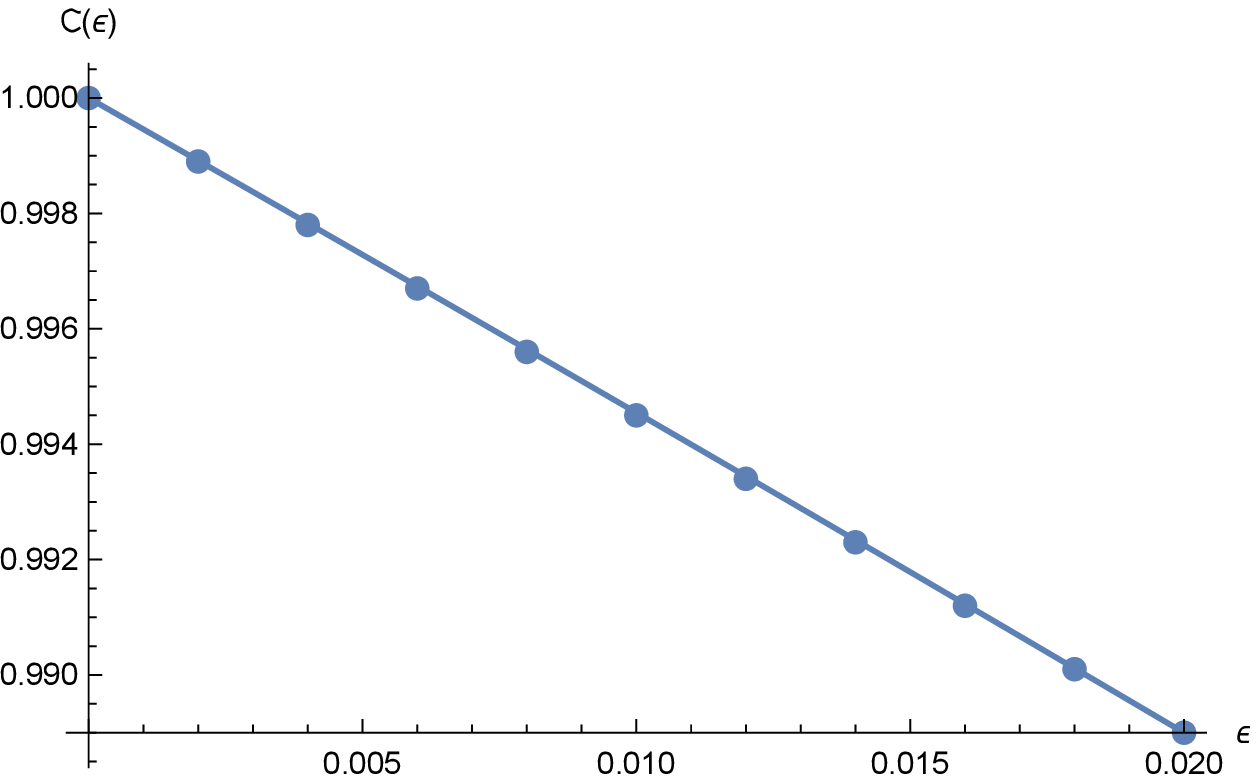}
\caption{The left hand graph shows the local slow roll correction factor 
$C(\epsilon)$ (solid blue), which was defined inexpression (\ref{Cdef}). Also
shown is its global approximation of $1 - \epsilon$ (dashed yellow) over the 
full inflationary range of $0 \leq \epsilon < 1$. The right hand graph shows 
$C(\epsilon)$ (solid blue) versus the better approximation of $1 - 0.55 \epsilon$ 
(large dots) relevant to the range $0 \leq \epsilon < 0.02$ favored by current 
data.}
\label{C(e)}
\end{figure}

\section{Simple Analytic Approximations} 

The exact analytic results of the previous section are valid for all single-scalar
models of inflation. However, they can be wonderfully simplified by exploiting the
fact that {\it the first slow roll parameter is very small.} The $95\%$ confidence 
bound on the tensor-to-scalar ration of $r < 0.12$ \cite{Ade:2015tva,Ade:2015xua} 
implies $\epsilon < 0.0075$. This suggests a number of approximations. First, the 
local slow roll correction factor $C(\epsilon_k)$, defined in (\ref{Cdef}), may as
well be set to unity. From Figure~\ref{C(e)} we see that the bound of $\epsilon < 
0.0075$ implies $1.0000 < C(\epsilon_k) < 0.9959$. This is not currently resolvable.

\begin{figure}[ht]
\includegraphics[width=6.0cm,height=4.8cm]{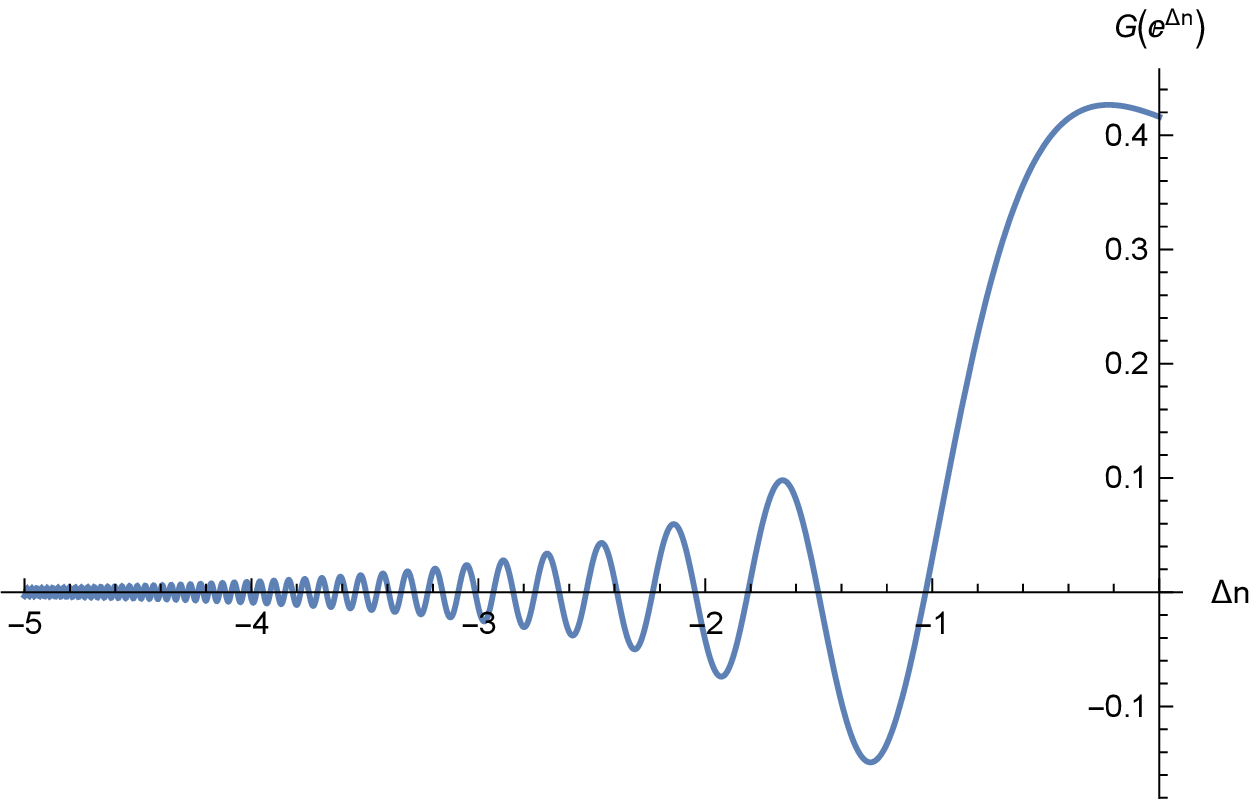}\hskip 1cm
\includegraphics[width=6.0cm,height=4.8cm]{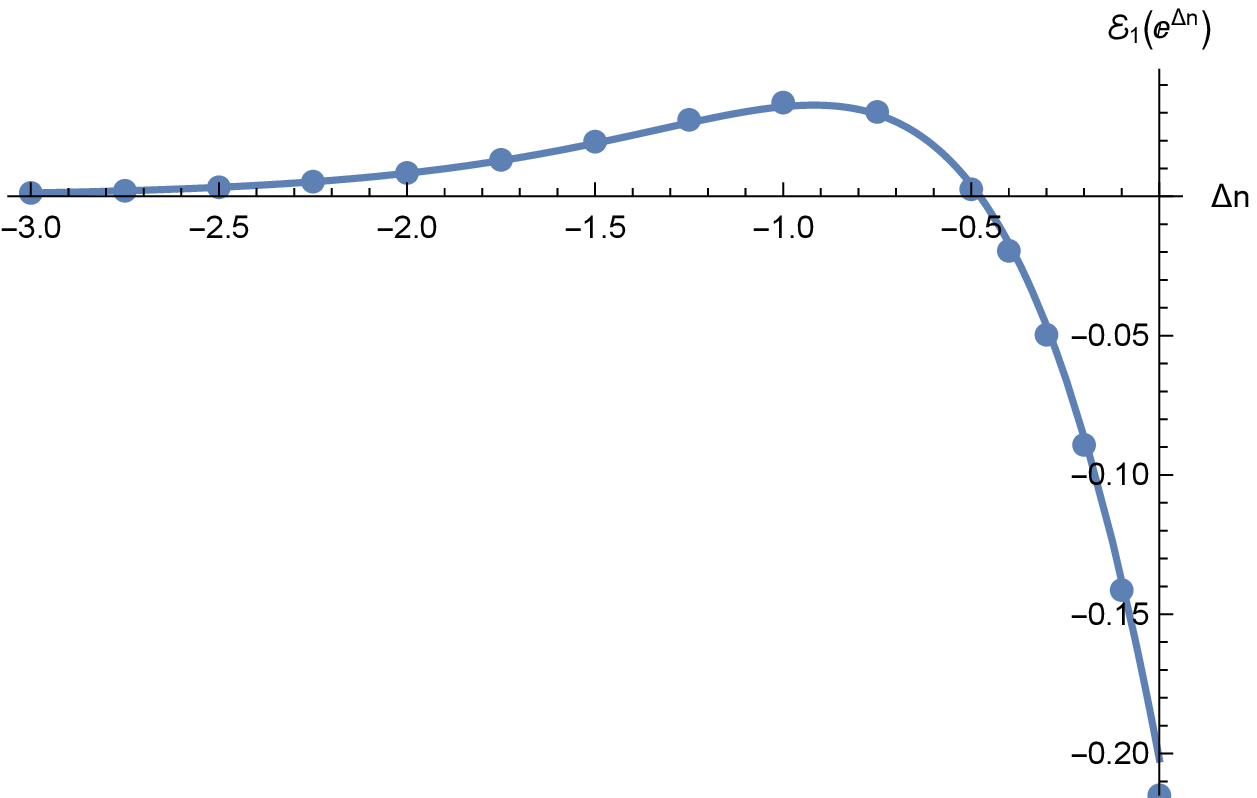}
\caption{The left hand graph shows the $\epsilon = 0$ Green's function
$G(e^{\Delta n})$ given in expression (\ref{Gdef}). The right hand graph
shows the coefficient of $\varepsilon''(n)$ in the small $\epsilon$ form
(\ref{Shsimple}) for $S_h(n,k)$. This function $\mathcal{E}_1(x)$ is 
defined by expressions (\ref{B0}), (\ref{A0}) and (\ref{E1}). The solid blue 
curve gives the exact numerical result while the large dots give the
approximation resulting from the series expansion on the right hand side of 
expression (\ref{A0}).}
\label{GE1}
\end{figure}

Another excellent approximation is taking $\epsilon = 0$ in the tensor and scalar 
Green's functions of expressions (\ref{tensorgreen}) and (\ref{scalargreen}),
\begin{eqnarray}
\lefteqn{\lim_{\epsilon = 0} G_h(n;m) = \lim_{\epsilon = 0} G_g(n;m) \equiv
G_0(n;m)} \nonumber \\
& & \hspace{1cm} = \frac{\theta(n \!-\! m)}{2} \, \Bigl[e^{\Delta m} \!+\! 
e^{3 \Delta m}\Bigr] \sin\Biggl[ -2 \Bigl\{ e^{-\Delta \ell} \!-\! 
{\rm arctan}\Bigl(e^{-\Delta \ell}\Bigr)\Bigr\} \Bigl\vert_{m}^{n} \Biggr] \; ,
\qquad \label{simplegreen}
\end{eqnarray}
where $\Delta m \equiv m - n_k$ and $\Delta \ell \equiv \ell - n_k$. Note that
this expression is valid before and after horizon crossing. An important special
case of (\ref{simplegreen}) is when $n$ becomes large, which gives the function
$G(e^{\Delta m})$ we define as,
\begin{equation}
G(x) \equiv \frac12 \Bigl(x \!+\! x^3\Bigr) \sin\Bigl[ \frac{2}{x} \!-\! 
2 \, {\rm arctan}\Bigl(\frac1{x}\Bigr) \Bigr] \; . \label{Gdef}
\end{equation}
From the graph in Figure~\ref{GE1} we see that $G(e^{\Delta n})$ suppresses
contributions more than a few e-foldings before horizon crossing.

\begin{figure}[ht]
\includegraphics[width=6.0cm,height=4.8cm]{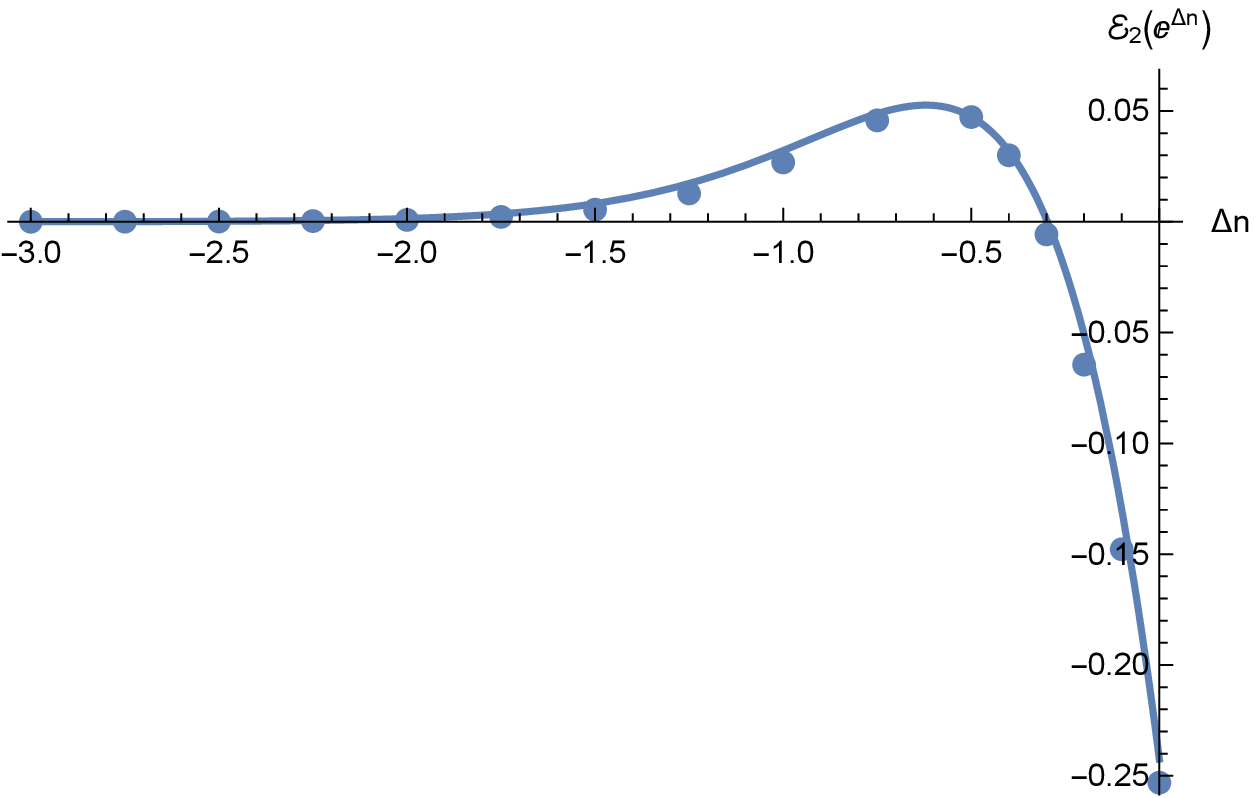}\hskip 1cm
\includegraphics[width=6.0cm,height=4.8cm]{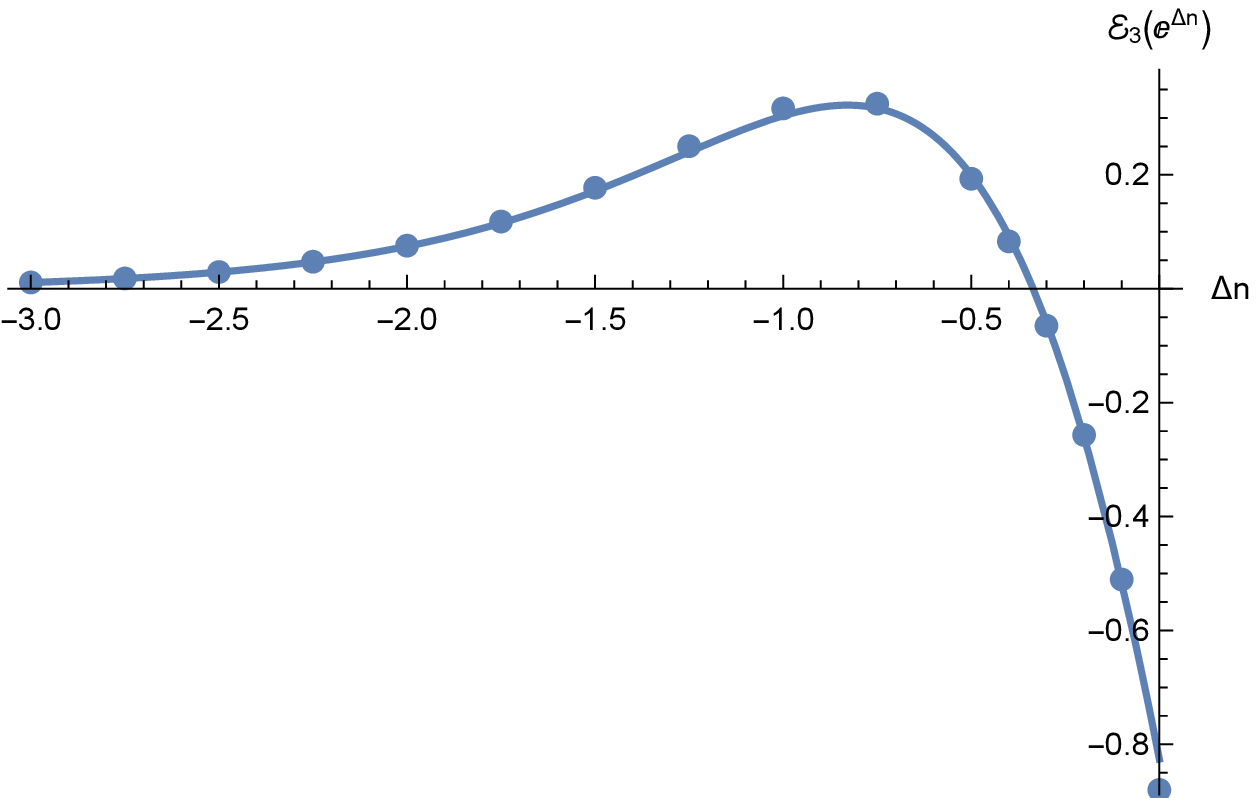}
\caption{The coefficients of $[\varepsilon'(n)]^2$ (left) and
$\varepsilon'(n)$ (right) in the small $\epsilon$ form (\ref{Shsimple})
for $S_h(n,k)$ . In each case the solid blue curve gives the exact 
numerical result, while the large dots give the result of using the 
series approximations on the far right of (\ref{A0}-\ref{D0}) in 
expressions (\ref{E2}) and (\ref{E3}).}
\label{E2E3}
\end{figure}

We can also take $\epsilon = 0$ in $\mathcal{H}$ and the derivatives of it 
in expressions (\ref{dH}) and (\ref{ddH}). This leads to exact results for 
$\mathcal{H}$, $\mathcal{B}$ and $\mathcal{E}$ in terms of the parameter 
$x \equiv e^{\Delta n}$,
\begin{eqnarray}
\lim_{\epsilon=0} \mathcal{H} & \equiv & \mathcal{H}_0(x) = x + x^3 \; , 
\label{H0} \\
\lim_{\epsilon = 0} \mathcal{B} & \equiv & \mathcal{B}_0(x) = 
\frac{-1 \!-\! 3 x^2}{1 \!+\! x^2} \; , \label{B0} \\
\lim_{\epsilon = 0} \mathcal{E} & \equiv & \mathcal{E}_0(x) = 
\frac{4 x^2}{(1 \!+\! x^2)^2} \; . \label{E0}
\end{eqnarray}
The three derivatives with respect to $\nu$ do not lead to simple expressions
even for $\epsilon \rightarrow 0$, but they can be well approximated over the 
range we require by short series expansions in powers of $x^2$,
\begin{eqnarray}
\lim_{\epsilon=0} \mathcal{A} & \equiv & \mathcal{A}_0(x) \simeq \frac{1.5 x^2 
\!+\! 1.8 x^4 \!-\! 1.5 x^6 \!+\! .63 x^8}{1 \!+\! x^2} \; , \qquad \label{A0} \\
\lim_{\epsilon=0} \, \mathcal{C} & \equiv & \mathcal{C}_0(x) \simeq \frac{x^2 \!+\! 
6.1 x^4 \!-\! 3.7 x^6 \!+\! 1.6 x^8}{(1 \!+\! x^2)^2} \; , \qquad \label{C0} \\
\lim_{\epsilon=0} \mathcal{D} & \equiv & \mathcal{D}_0(x) \simeq \frac{-3 x^2 
\!-\! 6.8 x^4 \!+\! 5.5 x^6 \!-\! 2.6 x^8}{(1 \!+\! x^2)^2} \; . \label{D0}
\end{eqnarray}
We can express the ratio of $\overline{H}/H$ in terms of the deviation
$\Delta \epsilon(n) \equiv \epsilon(n) - \epsilon_k$,
\begin{equation}
\frac{\overline{H}^2}{H^2} - 1 = \exp\Bigl[2 \!\! \int_{n_k}^n \!\!\! dm \,
\Delta \epsilon(m)\Bigr] -1 \simeq 2 \!\! \int_{n_k}^n \!\!\! dm \,
\Delta \epsilon(m) \; .
\end{equation}
All of this gives an approximation for the tensor source (\ref{Shcomplete}),
\begin{eqnarray}
\lefteqn{S_h(n,k) \simeq -2 \theta(-\Delta n) \Bigl[ \epsilon'' 
\mathcal{E}_1(e^{\Delta n}) \!+\! {\epsilon'}^2 \mathcal{E}_2(e^{\Delta n}) 
\!+\! \epsilon' \mathcal{E}_3(e^{\Delta n})\Bigr] + 2 \delta(\Delta n) \epsilon' 
\mathcal{E}_1(1) } \nonumber \\
& & \hspace{2cm} + 2 \theta(\Delta n) \Biggl\{ \Delta \epsilon(n) \!+\! 
\Bigl( \frac{4 \!+\! 2 e^{2 \Delta n}}{1 \!+\! e^{2 \Delta n}}\Bigr) \!\! 
\int_{n_k}^n \!\!\! dm \, \Delta \epsilon(m) \Biggr\} \frac{2}{1 \!+\! 
e^{2 \Delta n}} \; , \label{Shsimple} \qquad
\end{eqnarray}
where the three coefficient functions are,
\begin{eqnarray}
\mathcal{E}_1(x) & = & -1 - \mathcal{A}_0(x) - \mathcal{B}_0(x) \; , 
\label{E1} \\
\mathcal{E}_2(x) & = & \frac12 \!-\! \mathcal{A}_0(x) \!-\! \mathcal{C}_0(x) 
\!-\! 2 \mathcal{D}_0(x) \!-\! \mathcal{E}_0(x) \!-\! \frac12 \Bigl[ 2 \!+\! 
\mathcal{A}_0(x) \!+\! \mathcal{B}_0(x)\Bigr]^2 \; , \qquad \label{E2} \\
\mathcal{E}_3(x) & = & -1 \!+\! \mathcal{A}_0(x) \mathcal{B}_0(x) \!+\! 
\mathcal{B}_0^2(x) \!+\! 2 \mathcal{D}_0(x) \!+\! 2 \mathcal{E}_0(x) \; .
\label{E3}
\end{eqnarray}
Figures~\ref{GE1} and \ref{E2E3} show the various coefficient functions.

\begin{figure}[ht]
\includegraphics[width=6.0cm,height=4.8cm]{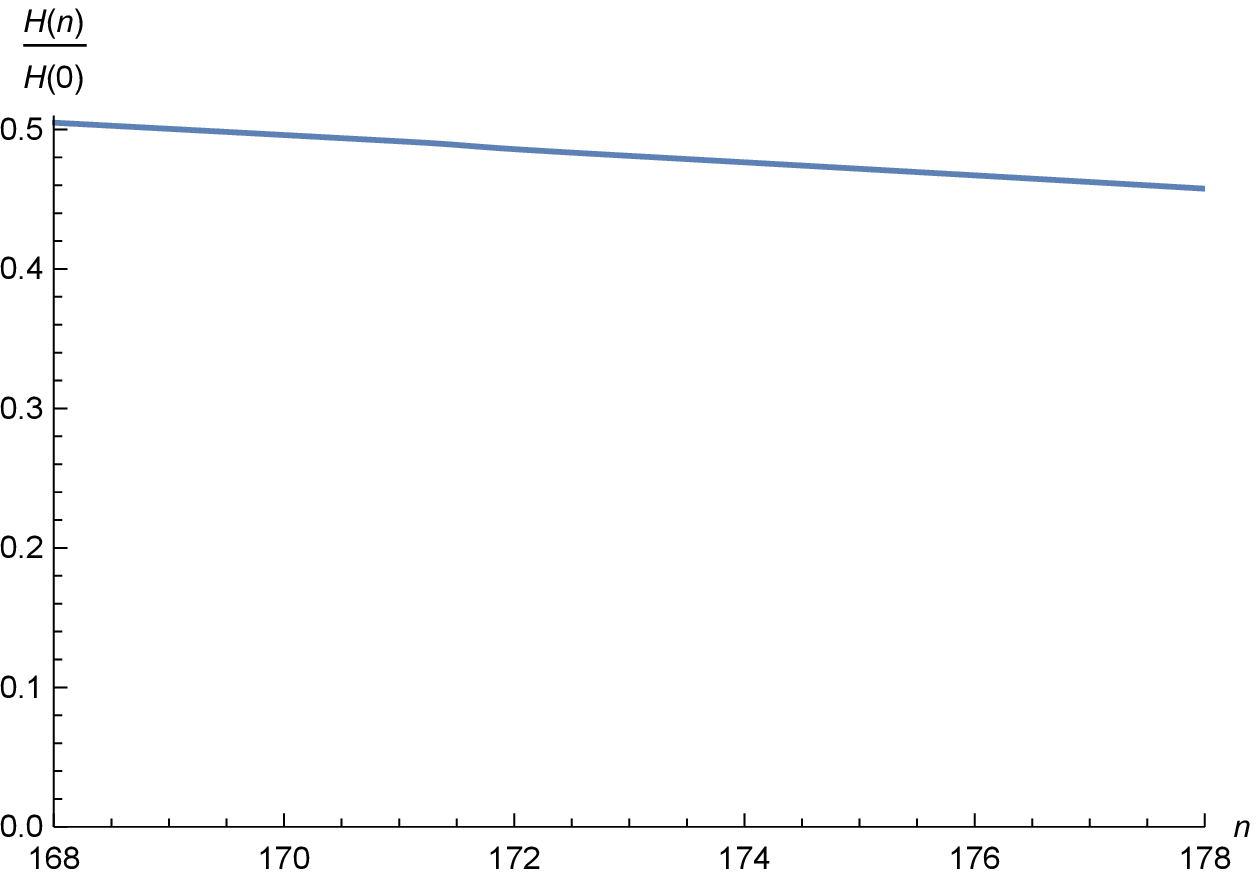}\hskip 1cm
\includegraphics[width=6.0cm,height=4.8cm]{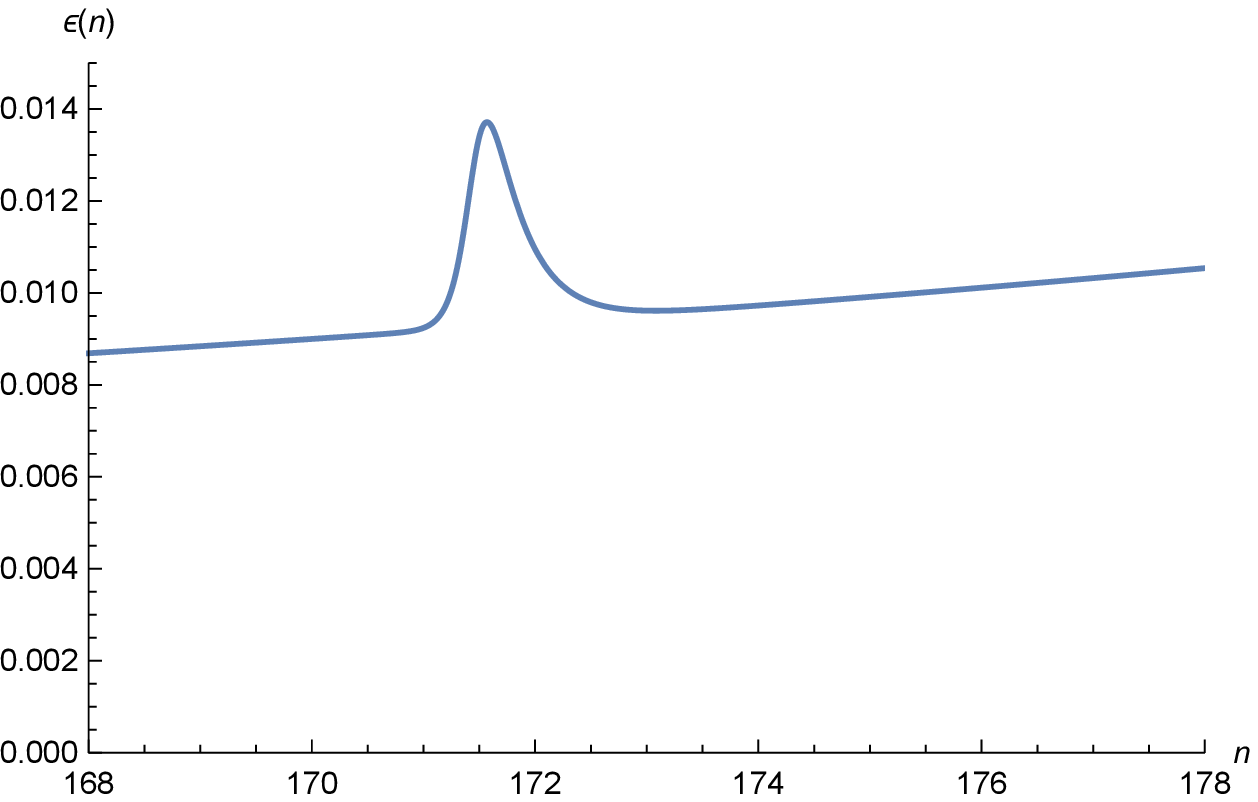}
\caption{The left hand figure shows the Hubble parameter and the
right shows the first slow roll parameter for a model with features.
This model which was proposed \cite{Adams:2001vc,Mortonson:2009qv} to 
explain the observed features in the scalar power spectrum at $\ell 
\approx 22$ and $\ell \approx 40$ which are visible in the data 
reported from both WMAP \cite{Covi:2006ci,Hamann:2007pa} and PLANCK 
\cite{Hazra:2014goa,Hazra:2016fkm}. Note that the feature has little
impact on $H(n)$ but it does lead to a distinct bump in $\epsilon(n)$.}
\label{stepgeometry}
\end{figure}

The smallness of $\epsilon$ means that the factors of $1/\epsilon$ which occur
in the scalar source (\ref{SgfromSh}) are hugely important. By comparison we 
can ignore the $S_h(n,k)$ terms and simply write,
\begin{equation}
S_g(n,k) \simeq -2\theta(-\Delta n) \Bigl[ \Bigl(\frac{\epsilon'}{\epsilon}\Bigr)'
\!+\! \frac12 \Bigl( \frac{\epsilon'}{\epsilon}\Bigr)^2 \!+\! 3 
\frac{\epsilon'}{\epsilon} \Bigr] + 2 \delta(\Delta n) \frac{\epsilon'}{\epsilon}
- 2\theta(\Delta n) \frac{\epsilon'}{\epsilon} \frac{2}{1 \!+\! e^{2 \Delta n}} 
\; . \label{Sgsimple}
\end{equation}
Because $\epsilon < 0.0075$ we expect $S_g$ to be more than 100 times 
as strong as $S_h$.

The approximations (\ref{simplegreen}), (\ref{Shsimple}) and (\ref{Sgsimple})
are valid so long as $\epsilon$ is small. If we additionally ignore nonlinear
terms in the equations for $h(n,k)$ and $g(n,k)$, the correction exponents
of expressions (\ref{fullDh}-\ref{fullDR}) become,
\begin{eqnarray}
\tau[\epsilon](k) & \simeq & \int_0^{n_k} \!\!\!\!\! dn \Biggl[ \epsilon''(n)
\mathcal{E}_1(e^{\Delta n}) \!+\! \Bigl[\epsilon'(n)\Bigr]^2
\mathcal{E}_2( e^{\Delta n}) \!\!+\! \epsilon'(n) \mathcal{E}_3(e^{\Delta n}) 
\Biggr] G(e^{\Delta n}) \nonumber \\
& & \hspace{-2cm} - \epsilon'(n_k) \mathcal{E}_1(1) G(1) -\!\! \int_{n_k}^{\infty}
\!\!\!\!\!\!\! dn \Biggl\{ \Delta \epsilon(n) \!+\! \Bigl(\frac{4 \!+\! 2 
e^{2 \Delta n}}{1 \!+\! e^{2 \Delta n}}\Bigr) \!\! \int_{n_k}^{n} \!\!\!\!\! dm 
\, \Delta \epsilon(m) \!\Biggr\} \frac{2 G(e^{\Delta n})}{1 \!+\! e^{2 \Delta n}} 
\; , \label{tensornonlocal} \\
\sigma[\epsilon](k) & \simeq & \int_{0}^{n_k} \!\!\!\!\! dn \Biggl[
\partial_n^2 \ln[\epsilon(n)] \!+\! \frac12 \Bigl( \partial_n 
\ln[\epsilon(n)]\Bigr)^2 \!+\! 3 \partial_n \ln[ \epsilon(n)]
\Biggr] G(e^{\Delta n}) \nonumber \\
& & \hspace{2cm} - \partial_{n_k} \ln[\epsilon(n_k)] \, G(1) +
\int_{n_k}^{\infty} \!\!\!\!\! dn \, \partial_n \ln[\epsilon(n)]
\frac{2 G(e^{\Delta n})}{1 \!+\! e^{2 \Delta n}} \; . 
\label{scalarnonlocal} \qquad
\end{eqnarray}
Recall that $\Delta n \equiv n - n_k$, $\Delta \epsilon(n) \equiv \epsilon(n)
- \epsilon_k$, the Green's function $G(e^{\Delta n})$ was defined in (\ref{Gdef}),
and the coefficient functions $\mathcal{E}_1(e^{\Delta n})$, 
$\mathcal{E}_2(e^{\Delta n})$ and $\mathcal{E}_3(e^{\Delta n})$ were given in
expressions (\ref{E1}-\ref{E3}). 

How large $\tau[\epsilon](k)$ and $\epsilon[\epsilon](k)$ are depends on what 
the inflationary model predicts for derivatives of $\epsilon(n)$. For example, 
the slow roll approximation of monomial inflation gives,
\begin{equation}
V(\varphi) = A \varphi^{\alpha} \qquad \Longrightarrow \qquad 
\epsilon(n) \simeq \frac{\epsilon_i}{1 \!-\! \frac{4}{\alpha} \epsilon_i n} 
\; . \label{monomialeps}
\end{equation}
For these models the various tensor and scalar contributions are small,
\begin{eqnarray}
V(\varphi) = A \varphi^{\alpha} \quad & \Longrightarrow & \quad 
\epsilon'' \simeq \frac{32}{\alpha^2} \, \epsilon^3 \;\; , \;\; 
{\epsilon'}^2 \simeq \frac{16}{\alpha^2} \, \epsilon^4 \;\; , \;\; 
\epsilon' \simeq \frac{4}{\alpha} \, \epsilon^2 \; , \label{monomialtensor} \\
& \Longrightarrow & \quad 
\Bigl(\frac{\epsilon'}{\epsilon}\Bigr)' \simeq \frac{16}{\alpha^2} \, 
\epsilon^2 \;\; , \;\; 
\Bigl(\frac{\epsilon'}{\epsilon}\Bigr)^2 \simeq \frac{16}{\alpha^2} \, 
\epsilon^2 \;\; , \;\; 
\frac{\epsilon'}{\epsilon} \simeq \frac{4}{\alpha} \, \epsilon \; . \qquad
\label{monomialscalar}
\end{eqnarray}

\begin{figure}[ht]
\includegraphics[width=6.0cm,height=4.8cm]{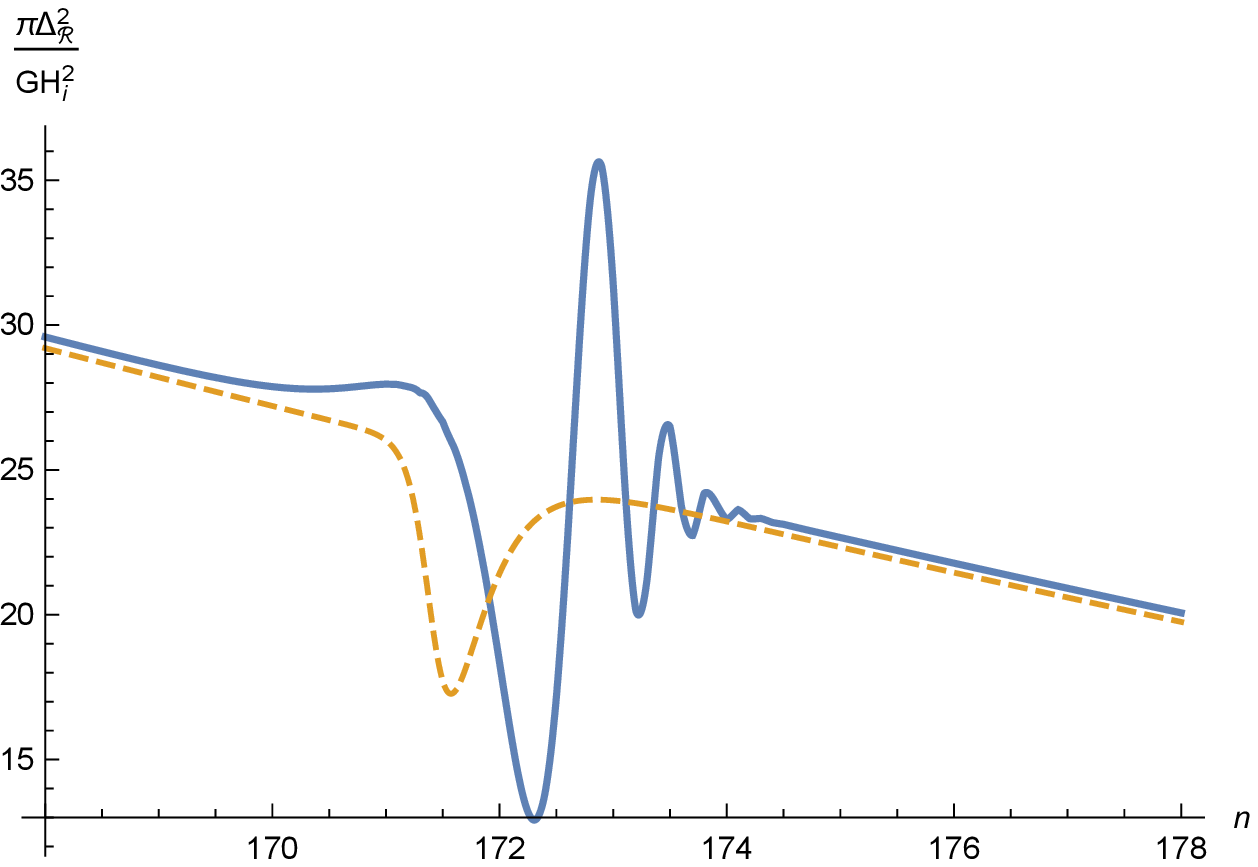}\hskip 1cm
\includegraphics[width=6.0cm,height=4.8cm]{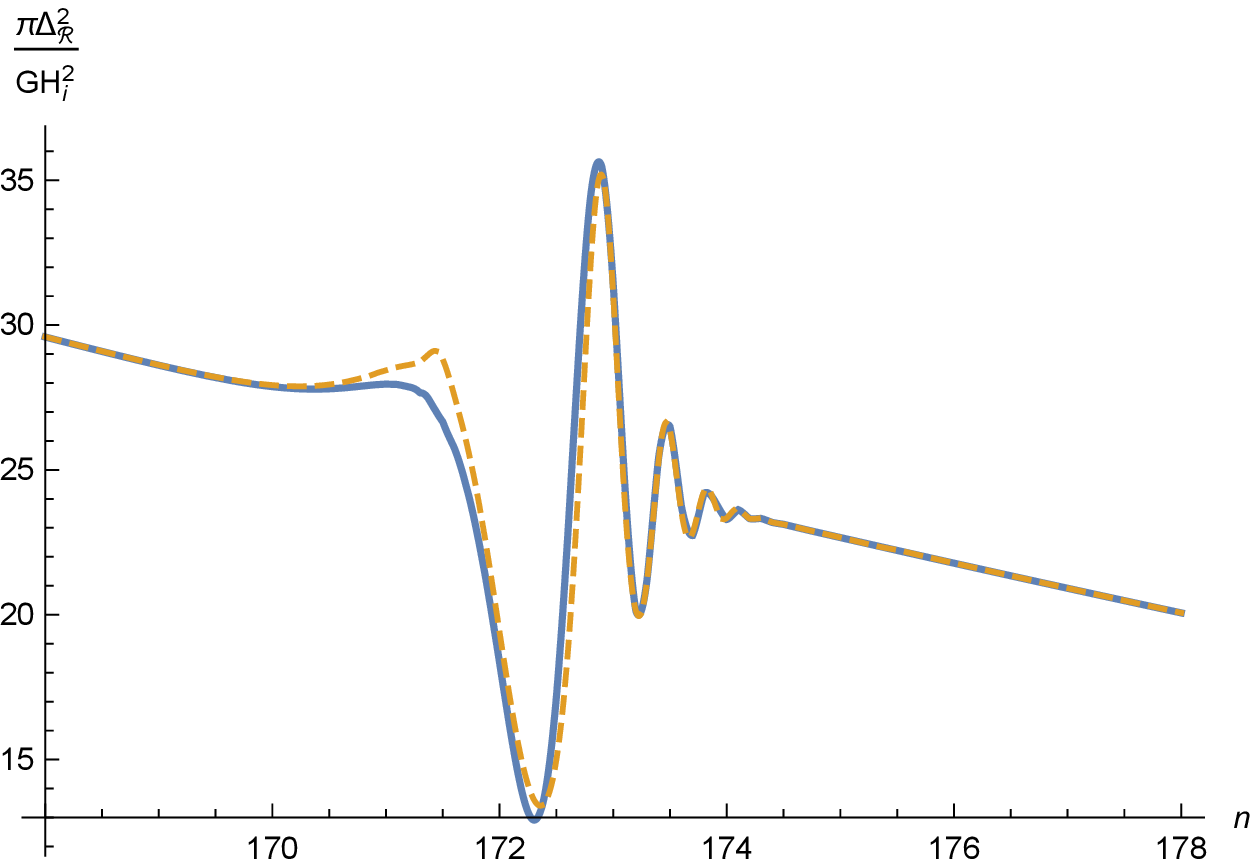}
\caption{These graphs show the scalar power spectrum for the model of 
Figure~\ref{stepgeometry}. The left hand figure compares the exact result
(solid blue) with the local slow roll approximation $\Delta^2_{\mathcal{R}}(k) 
\approx G H^2_k/\pi \epsilon_k \times C(\epsilon_k)$ (yellow dashed). The 
right hand figure compares the exact result (solid blue) with the much
better approximation (yellow dashed) obtained from multiplying by 
$\exp[\sigma[\epsilon](k)]$, using our analytic approximation 
(\ref{scalarnonlocal}) for $\sigma[\epsilon](k)]$.}
\label{approximation}
\end{figure}

The data disfavors monomial inflation \cite{Ade:2015tva,Ade:2015xua,Ade:2015lrj}, 
but $\tau[\epsilon](k)$ and $\sigma[\epsilon](k)$ will be small for any model 
which has only slow evolution of $\epsilon(n)$. Much larger effects occur for 
models with ``features'', which are transient fluctuations above or below the 
usual smooth fits \cite{Covi:2006ci}. Features imply short-lived changes in 
$\epsilon(n)$, which do not have much effect on $H(n)$ but can lead to large 
values of $\epsilon'(n)$ and $\epsilon''(n)$. Figure~\ref{stepgeometry} shows
$H(n)$ and $\epsilon(n)$ for a model that was proposed \cite{Adams:2001vc,
Mortonson:2009qv} to explain a deficit at $\ell \approx 22$, and an excess at 
$\ell \approx 40$, in the data reported by both WMAP \cite{Covi:2006ci,
Hamann:2007pa} and PLANCK \cite{Hazra:2014goa,Hazra:2016fkm}. In the range $171 
< n < 172.5$ the scalar experiences a step in its potential which has little 
effect on $H(n)$ but leads to a noticeable bump in $\epsilon(n)$.

Figure~\ref{approximation} shows shows the scalar power spectrum for the
model of Figure~\ref{stepgeometry}. The left hand graph compares the exact
result to the local slow roll approximation, without including the nonlocal 
conrrections from $\sigma[\epsilon](k)$. Not even the main feature is 
correct, and the secondary oscillations are completely absent. There is 
also a small systematic offset before and after the features. The right
hand graph shows the effect of adding $\sigma[\epsilon](k)$ with our
approximation (\ref{scalarnonlocal}). The agreement is almost perfect,
with the small remaining deviations attributable to nonlinear effects. The
small offset of the left hand graph (before and after the features) is due 
to the local slow roll approximation missing the steady growth which
$\epsilon(n)$ needs to reach the threshold of $\epsilon = 1$ at which 
inflation ends. We conclude:
\begin{enumerate}
\item{The nonlocal correction $\sigma[\epsilon](k)$ fixes the systematic 
under-prediction of the local slow roll approximation when $\epsilon(n)$ is
growing steadily;}
\item{The nonlocal correction $\sigma[\epsilon](k)$ makes large and essential
contributions when features are present; and}
\item{The nonlocal correction $\sigma[\epsilon](k)$ is well approximated by
(\ref{scalarnonlocal}).}
\end{enumerate}

\begin{figure}[ht]
\includegraphics[width=6.0cm,height=4.8cm]{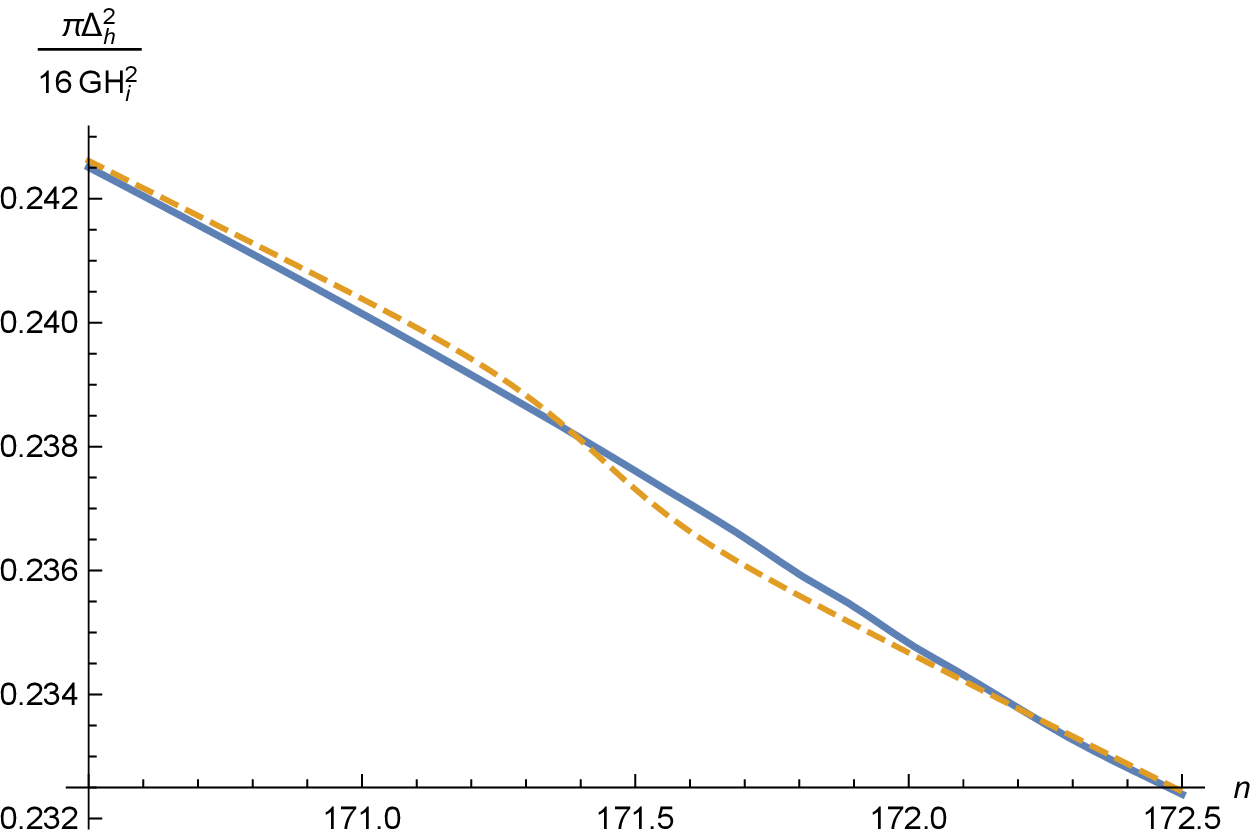}\hskip 1cm
\includegraphics[width=6.0cm,height=4.8cm]{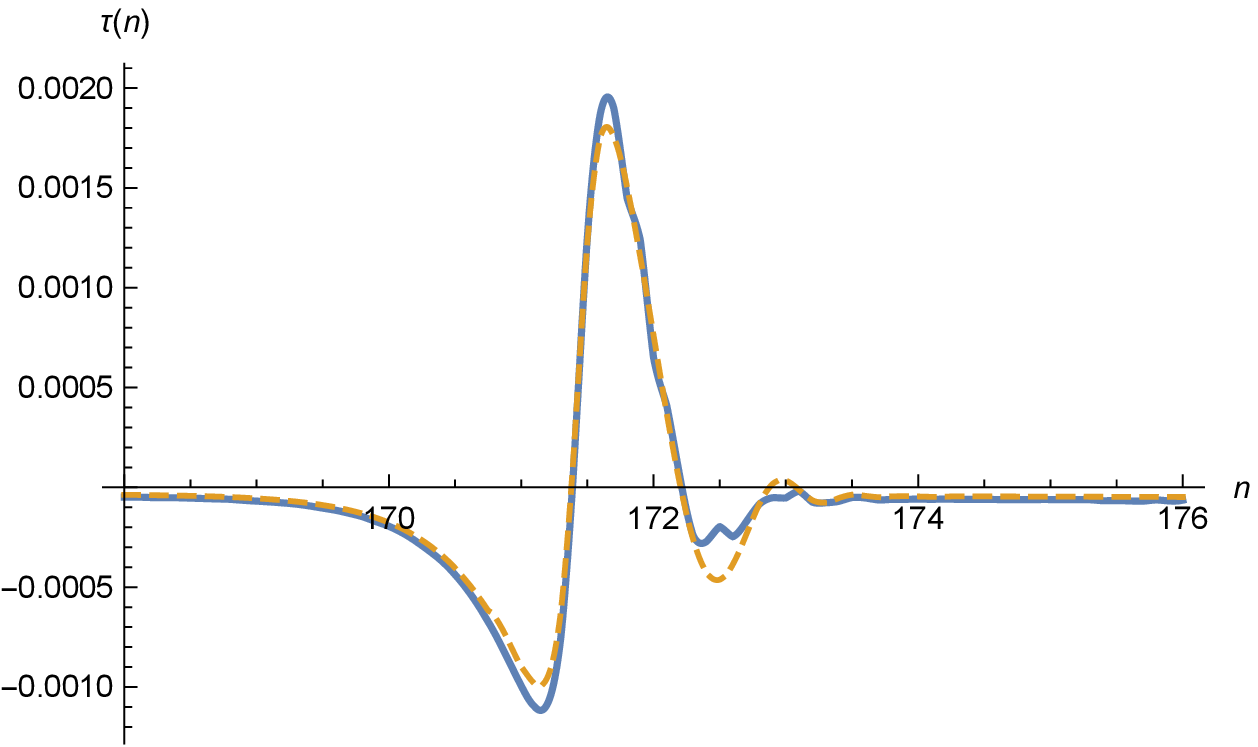}
\caption{These graphs show the tensor power spectrum for the model of
Figure~\ref{stepgeometry}. The left hand figure compares the exact result
(solid blue) with the local slow roll approximation $\Delta^2_{h}(k) \approx 
\frac{16}{\pi} G H_k^2 C(\epsilon_k)$ (yellow dashed). The solid blue line on 
the right hand graph shows the logarithm of the ratio of $\Delta^2_{h}(k)$ to
its local slow roll approximation. The yellow dashed line gives the nonlocal 
corrections of expression (\ref{tensornonlocal}).}
\label{StepTensor}
\end{figure}

Figure~\ref{StepTensor} shows the tensor power spectrum for the model of
Figure~\ref{stepgeometry}. The left hand graph compares the exact result
with the local slow roll approximation. The prominent features of the scalar 
power spectrum which can be seen in Figure~\ref{approximation} are several 
hundred times smaller, inverted and phase shifted, but they can just be made
out. The right hand graph compares our approximation (\ref{tensornonlocal})
for $\tau[\epsilon](k)$ with the exact result. The agreement is again almost 
perfect, with the small deviations actually attributable to numerical 
roughness in the interpolation of the exact computation, rather than to any
problem with our approximation (\ref{tensornonlocal}). Correlating tensor 
features with their much stronger scalar counterparts might be possible in 
the far future and would represent an impressive confirmation of single-scalar 
inflation \cite{Brooker:2016imi}.

\section{Reconstructing the Geometry}

We have so far considered the problem of using the inflationary geometry to
predict the power spectra. Here we wish to consider the inverse problem of
using $\Delta^2_{\mathcal{R}}(k)$ and $\Delta^2_{h}(k)$ to reconstruct $H(n)$
and $\epsilon(n)$. (The scalar and its potential can be derived from $H(n)$ 
and $\epsilon(n)$ by the formulae given in footnote 1.) It is well to begin by 
setting down a few general principles:
\begin{enumerate}
\item{Although $\Delta^2_{\mathcal{R}}(k)$ is measured to 3-digit accuracy,
the tensor power spectrum has yet to be resolved. When $\Delta^2_{h}(k)$ is
finally detected it will take a number of years before much precision is
attained. Therefore, reconstruction should be based on $\Delta^2_{\mathcal{R}}(k)$, 
with $\Delta^2_{h}(k)$ used only to fix the integration constant which gives 
the scale of inflation.}
\item{The first slow roll parameter is so small that there is no point in 
using the exact expression (\ref{fullDR}) for $\Delta^2_{\mathcal{R}}(k)$. 
Figure~\ref{C(e)} shows that we can ignore the local slow roll correction 
factor $C(\epsilon_k)$. Although the nonlocal correction exponent 
$\sigma[\epsilon](k)$ must be included, Figure~\ref{approximation} shows that 
the approximation (\ref{scalarnonlocal}) almost perfect.}
\item{The fact that $\epsilon(n)$ is small and smooth, with small transients,
motivates a hierarchy between $H$, $\epsilon$ and $\epsilon'/\epsilon$ based 
on calculus,
\begin{equation}
H(n) = H_i \exp\Bigl[-\!\! \int_0^n \!\!\! dm \, \epsilon(m)\Bigr] \;\; , \;\;
\epsilon(n) = \epsilon_i \exp\Bigl[ \int_0^n \!\!\! dm \, \frac{\epsilon'(m)}{
\epsilon(m)} \Bigr] \; . 
\end{equation}
Hence $H(n)$ is insensitive to small errors in $\epsilon(n)$, and $\epsilon(n)$
is insensitive to small errors in $\partial_n \ln[\epsilon(n)]$.}
\end{enumerate}

\begin{figure}[ht]
\includegraphics[width=6.0cm,height=4.8cm]{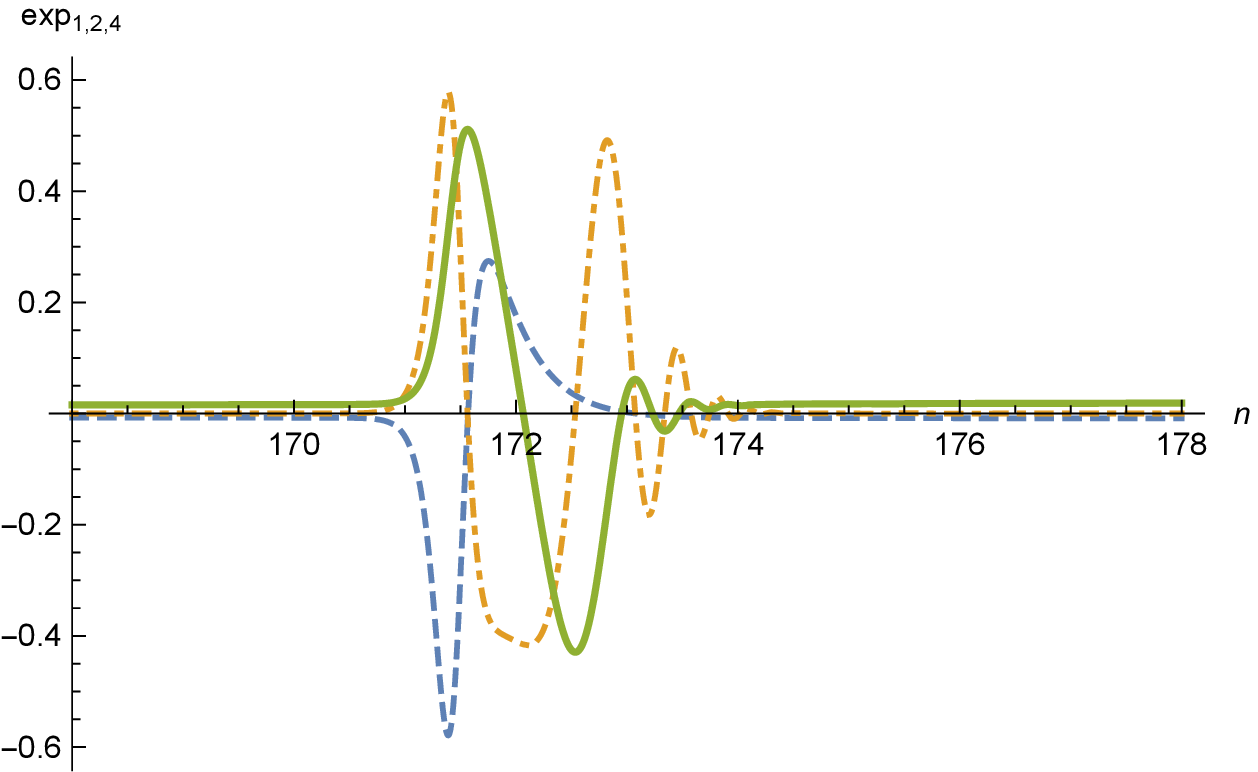}\hskip 1cm
\includegraphics[width=6.0cm,height=4.8cm]{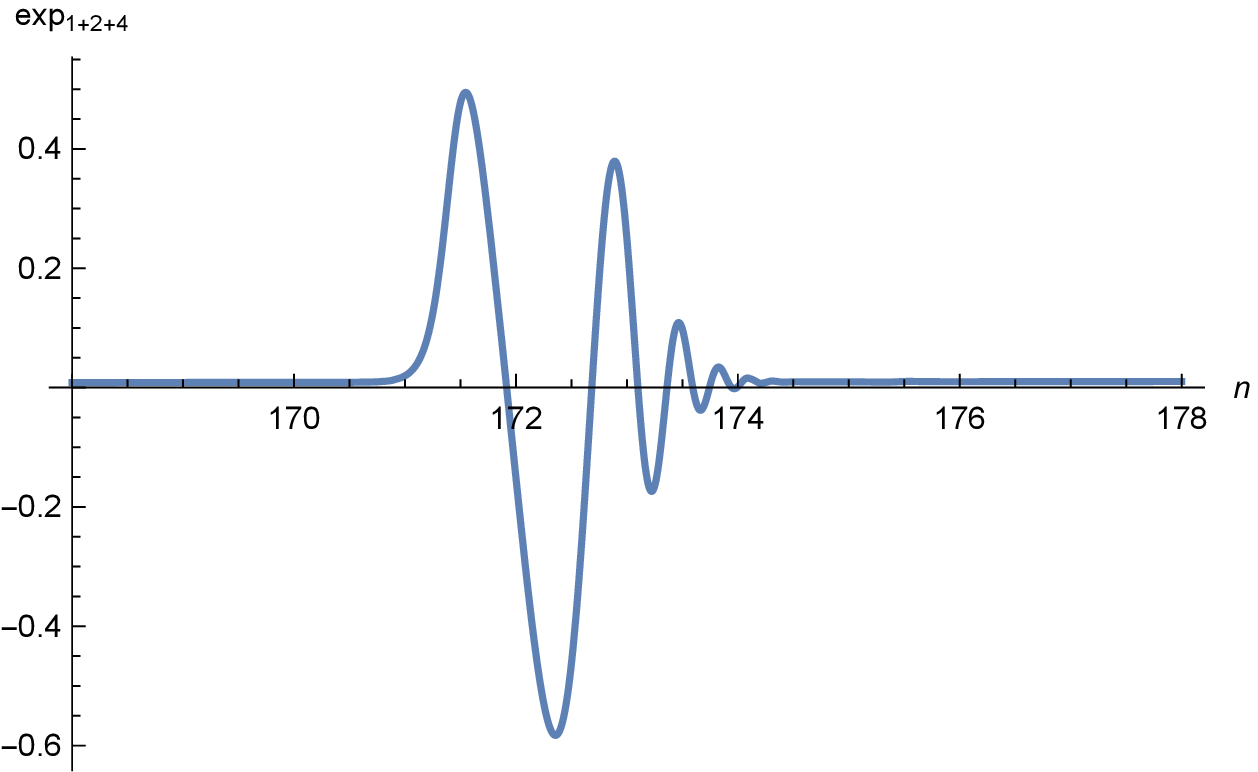}
\caption{Numerical values of exponents 1, 2 and 4 for the model of 
Figure~\ref{stepgeometry}. The left hand graph gives separate results for
expression (\ref{exp1}) in dashed blue, expression (\ref{exp2}) in dot-dashed
yellow, and expression (\ref{exp4}) in solid green. The right hand graph 
shows the sum of all three exponents.}
\label{exp124}
\end{figure}

We begin by converting from wave number $k$ to $n_k$, the number of e-foldings
since the beginning of inflation that $k$ experienced first horizon crossing.
It is also desirable to factor out the scale of inflation $H_i \equiv H(0)$,
\begin{equation}
h(n) \equiv \frac{H(n)}{H_i} \qquad , \qquad \delta(n_k) \equiv \frac{\pi
\Delta^2_{\mathcal{R}}(k)}{G H_i^2} \; .
\end{equation}
($H_i$ is the single number which would come from the tensor power spectrum.)
Based on the three principles we base reconstruction on the formula,
\begin{equation}
\delta(n) \simeq \frac{h^2(n)}{\epsilon(n)} \times \exp\Biggl[ \sum_{i=1}^5
{\rm exp}_i(n)\Biggr] \; , \label{reconeqn1}
\end{equation}
where the five exponents follow from our approximation (\ref{scalarnonlocal})
for $\sigma[\epsilon](k)$, 
\begin{eqnarray}
{\rm exp}_1(n) & = & -\partial_{n} \ln[\epsilon(n)] \!\times\! G(1) \; , 
\label{exp1} \\
{\rm exp}_2(n) & = & \int_0^{n} \!\!\! dm \, \partial_{m}^2 \ln[\epsilon(m)] 
\!\times\! G(e^{m-n}) \; , \label{exp2} \\
{\rm exp}_3(n) & = & \frac12 \! \int_0^{n} \!\!\! dm \, \Bigl[\partial_m 
\ln[\epsilon(m)]\Bigr]^2 \!\times\! G( e^{m-n}) \; , \label{exp3} \\
{\rm exp}_4(n) & = & 3\! \int_0^{n} \!\!\! dm \, \partial_{m} \ln[\epsilon(m)] 
\!\times\! G(e^{m-n}) \; , \label{exp4} \\
{\rm exp}_5(n) & = & 2\! \int_{n}^{\infty} \!\! dm \, \partial_m \ln[\epsilon(m)] 
\!\times\! \frac{G(e^{m-n})}{1 \!+\! e^{2(m-n)}} \; . \label{exp5}
\end{eqnarray}

\begin{figure}[ht]
\includegraphics[width=6.0cm,height=4.8cm]{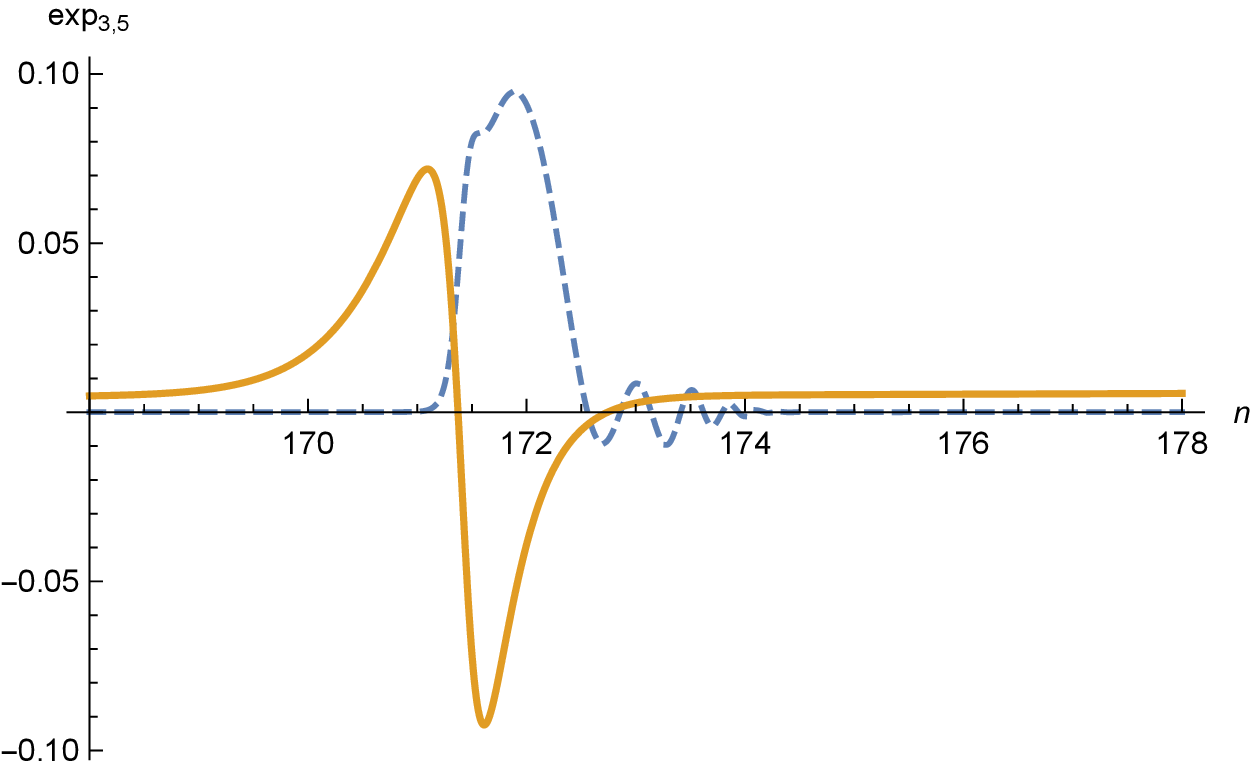}\hskip 1cm
\includegraphics[width=6.0cm,height=4.8cm]{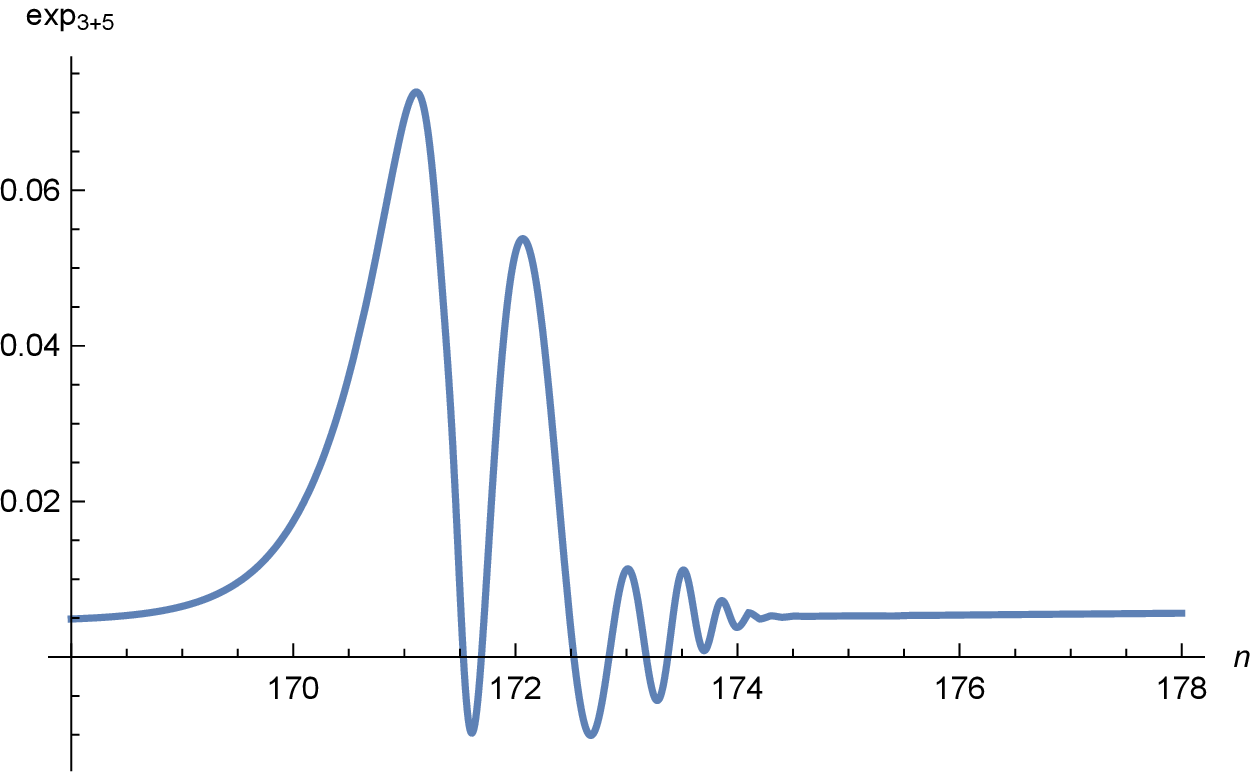}
\caption{Numerical values of ${\rm exp}_3(n)$ and ${\rm exp}_5(n)$ for 
the model of Figure~\ref{stepgeometry}. The left hand graph gives 
separate results for expression (\ref{exp3}) in dashed blue, and expression 
(\ref{exp5}) in solid yellow. Note that ${\rm exp}_5(n)$ is responsible for 
correcting the small, systematic under-prediction of the slow roll 
approximation before and after the feature. The right hand graph shows the 
sum.}
\label{exp35}
\end{figure}

To just reconstruct the Hubble parameter there is no need to include the 
correction exponents (\ref{exp1}-\ref{exp5}). Using only the leading slow 
roll terms gives,
\begin{equation}
\delta(n) \simeq \frac{h^2(n)}{\epsilon(n)} \quad \Longrightarrow \quad
h^2(n) \simeq \frac{1}{1 \!+\! \int_{0}^{n} \! \frac{2 dm}{\delta(m)}}
\; . \label{slowrollH}
\end{equation}
Even for the power spectrum of Figure~\ref{approximation} the reconstruction
of $h(n)$ given by expression (\ref{slowrollH}) is barely distinguishable
from the left hand graph of figure~\ref{stepgeometry}.

\begin{figure}[ht]
\includegraphics[width=6.0cm,height=4.8cm]{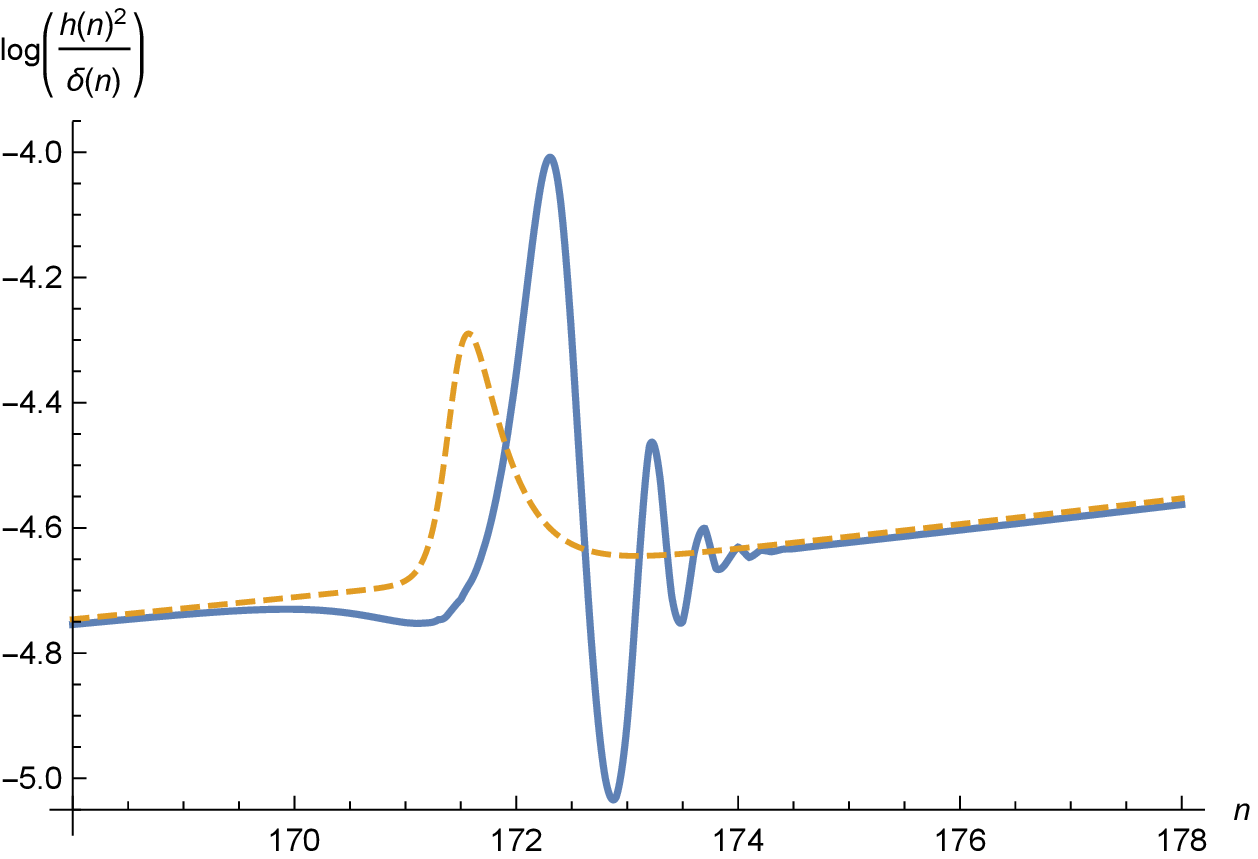}\hskip 1cm
\includegraphics[width=6.0cm,height=4.8cm]{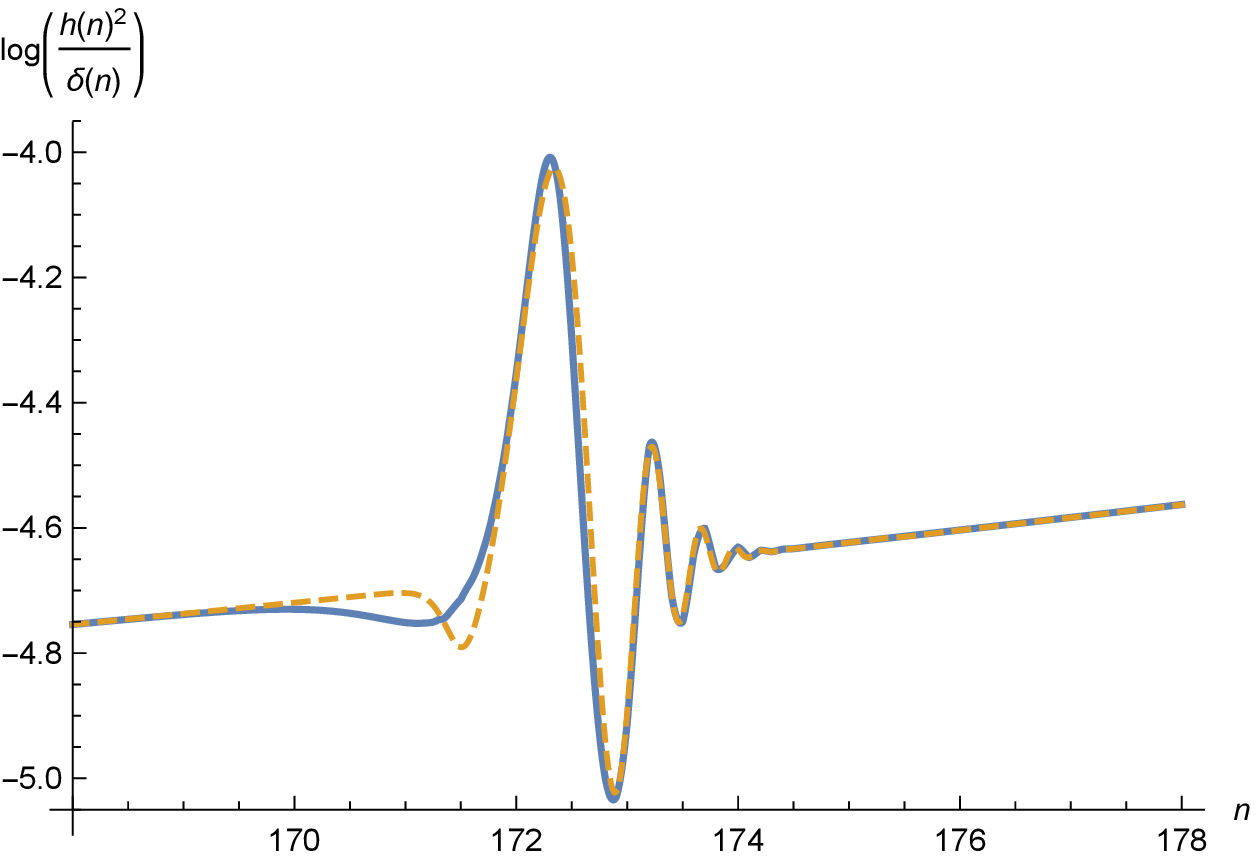}
\caption{Various choices for the left hand side of the first pass
reconstruction equation for the model of Figure~\ref{stepgeometry}. The 
left hand graph shows the first pass source $-\ln[\delta(n)] + 2 \ln[h(n)]$ 
in solid blue with $\ln[\varepsilon(n)]$ overlaid in dashed yellow. The poor 
agreement between the two curves is why using just $\ln[\varepsilon(n)]$ as
the left hand side of the first pass reconstruction fails to converge when
features are present. The right hand graph shows the much better agreement 
between the same source (solid blue) and $\ln[\varepsilon(n)] - {\rm exp}_1(n) 
- {\rm exp}_2(n) - {\rm exp}_4(n)$ (dashed yellow).}
\label{fits}
\end{figure}

Not all the exponents (\ref{exp1}-\ref{exp5}) are equally important.
Figures~\ref{exp124} and \ref{exp35} show that the set of ${\rm exp}_1(n)$, 
${\rm exp}_2(n)$ and ${\rm exp}_4(n)$ are about ten times larger than 
${\rm exp}_3(n)$ and ${\rm exp}_5(n)$ for the model of Figure~\ref{stepgeometry}.
That reconstructing features indeed requires the three large exponents is
apparent from Figure~\ref{fits}. Taking the logarithm of (\ref{reconeqn1}) and
moving the three large exponents to the left gives,
\begin{eqnarray}
\lefteqn{\Bigl[1 \!+\! G(1) \partial_n \Bigr] \ln[\epsilon(n)] - \int_0^{n} 
\!\! dm \Bigl[ \partial_m^2 \!+\! 3 \partial_m\Bigr] \ln[\epsilon(m)] 
\!\times\! G(e^{m-n}) } \nonumber \\
& & \hspace{3.5cm} \simeq -\ln[\delta(n)] + 2 \ln[h(n)] + {\rm exp}_3(n) +
{\rm exp}_5(n) \; . \qquad \label{reconeqn2}
\end{eqnarray}
This becomes a linear, nonlocal equation for $\ln[\epsilon(n)]$ if we drop
${\rm exp}_3(n)$ and ${\rm exp}_5(n)$ and use expression (\ref{slowrollH}) 
for the Hubble parameter,
\begin{eqnarray}
\lefteqn{\Bigl[1 \!+\! G(1) \partial_n \Bigr] \ln[\epsilon(n)] - \int_0^{n} 
\!\! dm \Bigl[ \partial_m^2 \!+\! 3 \partial_m\Bigr] \ln[\epsilon(m)] 
\!\times\! G(e^{m-n}) } \nonumber \\
& & \hspace{6cm} \simeq -\ln[\delta(n)] - \ln\Biggl[1 \!+\! \int_{0}^{n} \!\!
\frac{2 dm}{\delta(m)} \Biggr] \; . \qquad \label{reconeqn3}
\end{eqnarray}

The linearity of equation (\ref{reconeqn3}) means that it can be solved by
a Green's function, in spite of being nonlocal. The required Green's function
becomes a symmetric function of its arguments if we note from Figure~\ref{GE1} 
and expression (\ref{Gdef}) that $G(e^{n-n_k})$ is essentially zero more than 
about $N \sim 4$ e-foldings before horizon crossing. The Green's function 
equation is,
\begin{equation}
\Bigl[1 \!+\! G(1) \partial_n\Bigr] \mathcal{G}(n) - \int_{-N}^{n} \!\!\! dm \,
(\partial_m^2 \!+\! 3 \partial_m) \mathcal{G}(m) \!\times\! G(e^{m-n}) =
\delta(n) \; . \label{keygreen}
\end{equation} 
We can solve (\ref{reconeqn3}) by integrating against the source on the right
hand side,
\begin{equation}
\ln[\epsilon(n)] = \int_0^{\infty} \!\! dm \, \mathcal{G}(n \!-\! m) \!\times\!
{\rm Source}(m) \; . \label{greensol}
\end{equation}
This might be regarded as the first pass of an iterative solution to 
(\ref{reconeqn2}). After the first pass solution of (\ref{reconeqn3}) one would 
use the resulting $\ln[\epsilon(n)]$ to construct $h(n)$ and to evaluate 
${\rm exp}_3(n)$ and ${\rm exp}_5(n)$ on the right hand side of (\ref{reconeqn2}). 
Then the same Green's function solution (\ref{greensol}) could be used with this
more accurate source to find a more accurate $\ln[\epsilon(n)]$, which would lead 
to a more accurate source, and so on.

\begin{figure}[ht]
\includegraphics[width=6.0cm,height=4.8cm]{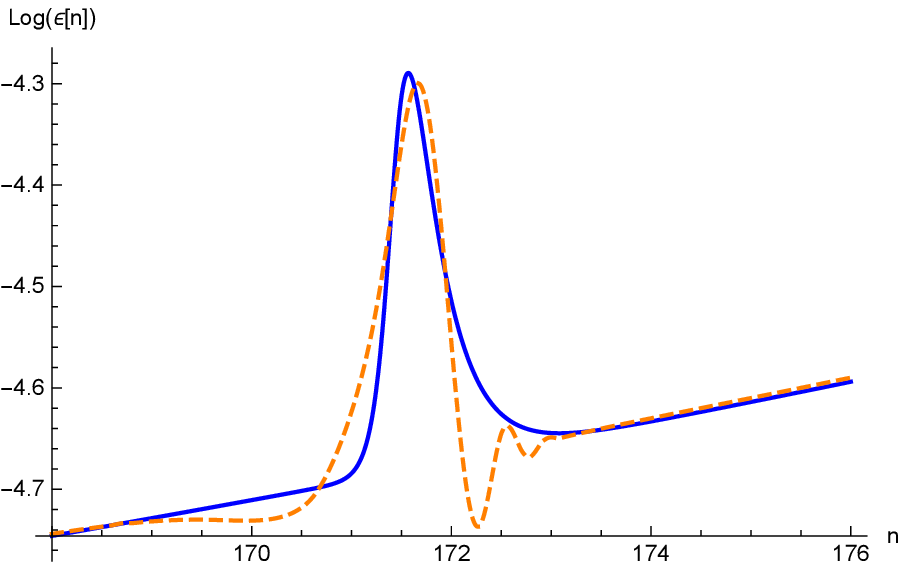}\hskip 1cm
\includegraphics[width=6.0cm,height=4.8cm]{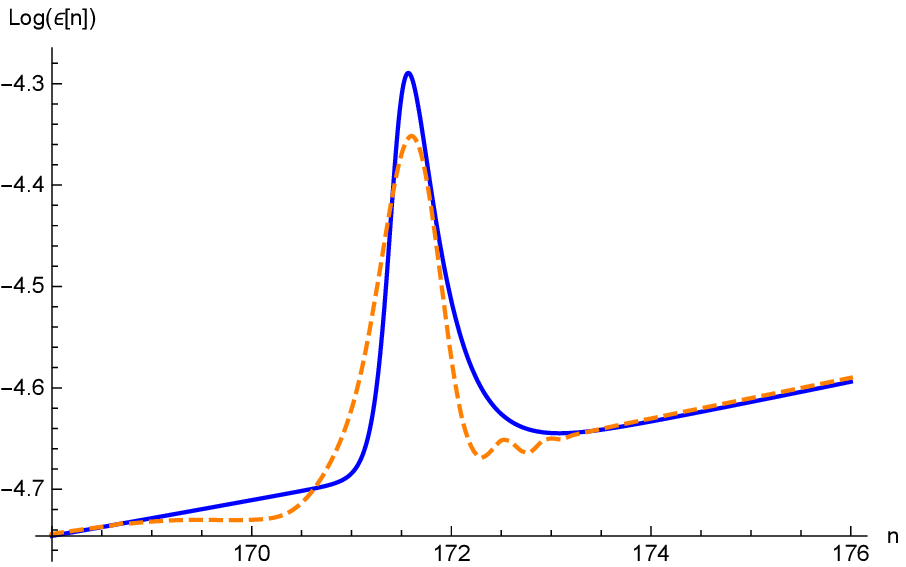}
\caption{These graphs show numerical reconstructions of $\ln[\epsilon(n)]$ for 
the power spectrum of Figure~\ref{approximation}. The solid blue line of the left 
hand graph shows the exact result while the yellow dashed line gives the result
of integrating $\mathcal{G}_0(n\!-\!m)$ --- using the first six terms of the sum 
over $\ell$ in expression (\ref{G0sol}) --- against the first pass source on the 
right hand side of (\ref{reconeqn3}). The right hand graph shows the result of 
adding the first order improvement $\mathcal{G}_1(n \!-\! m)$ --- computed using 
the first four terms of the sum over $m$ in expression (\ref{expI1}).}
\label{G0G1}
\end{figure}

We are not able to solve (\ref{keygreen}) exactly owing to the factor of
$G(e^{m-n})$ inside the integral. Consideration of Figure~\ref{GE1} suggests
that this troublesome factor might be approximated as a square wave of width
$\Delta = 0.8$,
\begin{equation}
G(e^{m-n}) \approx G(1) \theta(n \!-\! m \!-\! \Delta) \; . \label{Gapprox}
\end{equation}
Making the approximation (\ref{Gapprox}) leads to a still-nonlocal equation,
\begin{equation}
(\partial_n \!+\! 3) \mathcal{G}_0(n \!-\! \Delta) - \alpha \mathcal{G}_0(n) = 
\frac{\delta(n)}{G(1)} \qquad , \qquad \alpha \equiv 3 - \frac1{G(1)} \; .
\label{G0eqn}
\end{equation}
The ``retarded'' solution to (\ref{G0eqn}) which avoids exponentially growing 
terms is,
\begin{equation}
\mathcal{G}_0(n) = \frac{e^{3 (n+\Delta)}}{G(1)} \sum_{\ell=0}^{\infty} 
\frac1{\ell!} \Bigl[ \alpha e^{-3\Delta} \Bigl(n \!+\! (\ell \!+\! 1) \Delta
\Bigr)\Bigr]^{\ell} \theta\Bigl(n \!+\! (\ell \!+\! 1) \Delta \Bigr) \; .
\label{G0sol}
\end{equation}
Figure~\ref{G0G1} shows the result of using just $\mathcal{G}_0(n)$ to 
reconstruct $\ln[\epsilon(n)]$ with the source taken as the right hand side 
of (\ref{reconeqn3}).

\begin{figure}[ht]
\includegraphics[width=6.0cm,height=4.8cm]{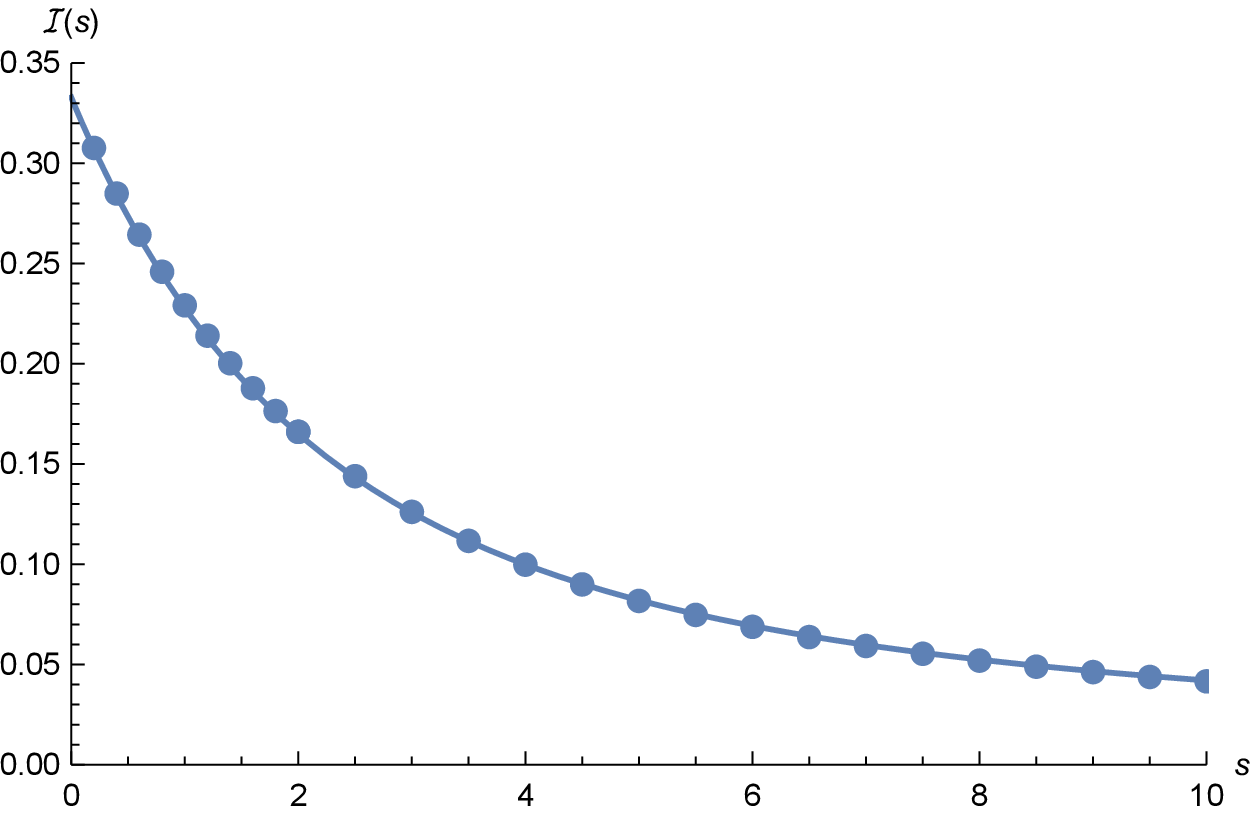}\hskip 1cm
\includegraphics[width=6.0cm,height=4.8cm]{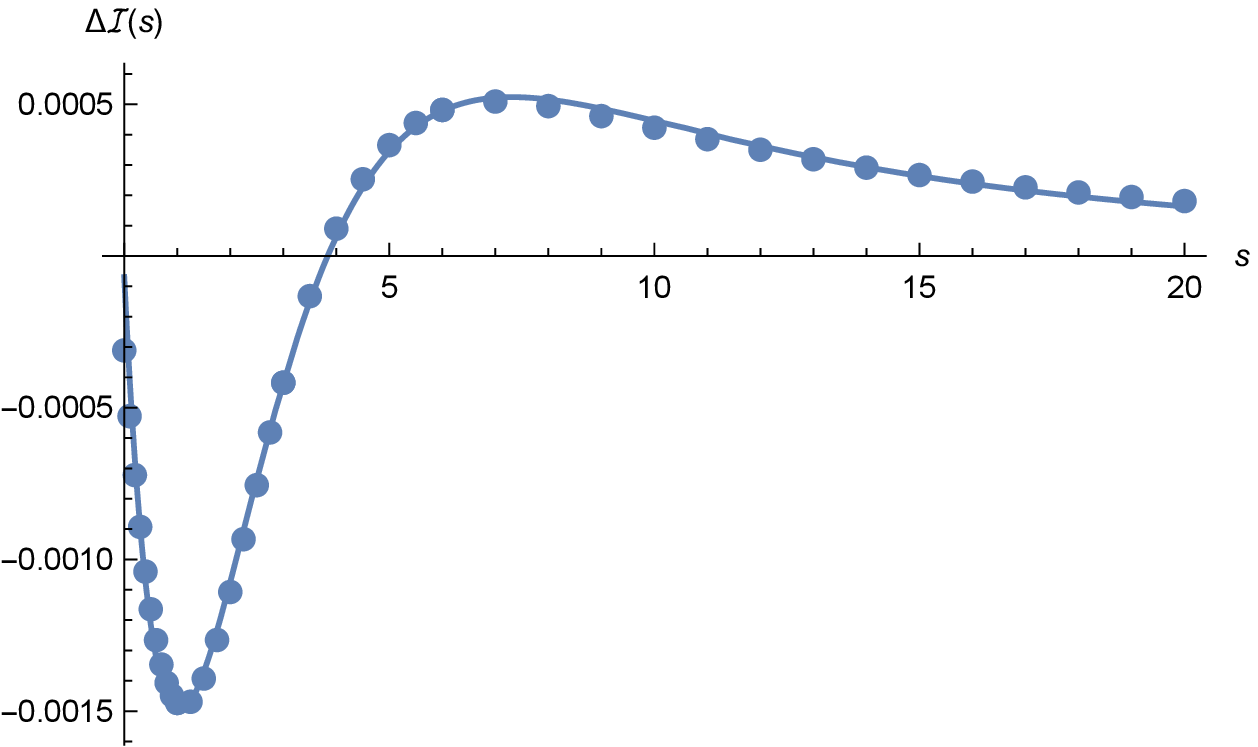}
\caption{The solid blue line of the left hand graph shows a numerical evaluation 
of the integral $\mathcal{I}(s)$ of expression (\ref{INT}). The 0th order
approximation $\mathcal{I}_0(s)$ of expression (\ref{INT0}) is overlaid in large 
dots. The solid blue line of the right hand graph shows the deviation $\Delta 
\mathcal{I}(s) \equiv \mathcal{I}(s) - \mathcal{I}_0(s)$. Our fit $\mathcal{I}_1(s)$
of expression (\ref{INT1}) is overlaid in large dots.}
\label{Laplace}
\end{figure}

Further improvement requires a better approximation for the Green's function
$\mathcal{G}(n)$. It is instructive to take the Laplace transform, restoring
the second argument of the Green's function,
\begin{equation}
\widehat{\mathcal{G}}(s;m) \equiv \int_0^{\infty} \!\! dn \, e^{-s n}
\mathcal{G}(n \!-\! m) \; .
\end{equation}
The Laplace transform of the Green's function equation (\ref{keygreen}) is,
\begin{equation}
\Bigl[1 \!+\! G(1) s \!-\! (s \!+\! 3) s \!\times\! \mathcal{I}(s)\Bigr]
\widehat{\mathcal{G}}(s;m) = e^{-m s} \; , \label{Laplacegreen}
\end{equation}
where we define, 
\begin{equation}
\mathcal{I}(s) \equiv \int_{0}^{\infty} \!\! d\ell \, e^{-s \ell} \!\times\!
G(e^{-\ell}) \; . \label{INT}
\end{equation}
The problem of approximating $\mathcal{G}(n-m)$ is therefore related to the
one of approximating (\ref{INT}), and of recognizing the resulting inverse
Laplace transform of $\widehat{\mathcal{G}}(s;m)$. Making the approximation
(\ref{Gapprox}) in (\ref{INT}) gives,
\begin{equation}
\mathcal{I}_0(s) = \frac{G(1)}{s} \Bigl[1 - e^{-0.8 s}\Bigr] \; . \label{INT0}
\end{equation}
Figure~\ref{Laplace} reveals that this is indeed a good approximation. 
Figure~\ref{Laplace} also shows that the small residual is well fit by the 
function,
\begin{equation}
\mathcal{I}_1(s) = \frac{0.154}{(s + 8.97)^2} \, \sin\Biggl[1.76 \Bigl(
1 \!-\! e^{-0.262 (s - 3.78)}\Bigr)\Biggr] \; . \label{INT1}
\end{equation}

To obtain the first correction to $\mathcal{G}_0(n - m)$  we begin by expanding 
$\widehat{\mathcal{G}}(s;m)$ in powers of $\mathcal{I}_1(s)$,
\begin{eqnarray}
\widehat{\mathcal{G}}(s;m) & \simeq & \frac{e^{-ms}}{G(1)[(s \!+\! 3) e^{-\Delta s}
\!-\! \alpha] \!-\! s (s \!+\! 3) \mathcal{I}_1(s)} \; , \\
& = & \frac{e^{-ms}}{G(1) [ (s \!+\! 3) e^{-\Delta s} \!-\! \alpha]} +
\frac{s (s \!+\! 3) \mathcal{I}_1(s) e^{-m s}}{G^2(1) [ (s \!+\! 3) e^{-\Delta s}
\!-\! \alpha]^2} + \dots \; , \qquad \\
& \equiv & \widehat{\mathcal{G}}_0(s;m) + \widehat{\mathcal{G}}_1(s;m) + \dots 
\end{eqnarray} 
We can recognize the inverse Laplace transform by expanding $\mathcal{I}_1(s)$,
\begin{eqnarray}
\lefteqn{\frac{a}{(s \!+\! e)^2} \sin\Bigl[b \!-\! b e^{-c (s + d)}\Bigr] = 
\frac{a}{(s \!+\! e)^2} \Biggl\{ \sin(b) \sum_{m=0}^{\infty} 
\frac{(-1)^m}{(2 m)!} \Bigl[ b e^{-c (s+d)} \Bigr]^{2m} } \nonumber \\
& & \hspace{5cm} - \cos(b) \sum_{m=0}^{\infty} \frac{(-1)^m}{(2 m \!+\! 1)!} 
\Bigl[b e^{-c (s + d)}\Bigr]^{2m + 1} \Biggr\} \; . \qquad \label{expI1}
\end{eqnarray}
Figure~\ref{G0G1} shows the effect of using $\mathcal{G}_0(n - m) + 
\mathcal{G}_1(n - m)$ to solve equation (\ref{reconeqn3}) approximately for
$\ln[\epsilon(n)]$.

Figure~\ref{G0G1} shows that additional improvements are needed before our
technique gives good results for $\partial_n \ln[\epsilon(n)]$ when features
are present. However, our results for $\epsilon(n)$ are already reasonable, 
and those for $h(n)$ are staggeringly accurate. For the model of 
Figure~\ref{stepgeometry} the largest percentage error on in reconstructing 
$\epsilon(n)$ is $2.2\%$, and the percentage error for $h(n)$ never exceeds 
$0.04\%$. This seems considerably better than the General Slow Roll 
Approximation \cite{Kadota:2005hv}, or techniques based on local expressions 
\cite{Barrow:2016wiy}. A recent proposal based on inverse-scattering 
\cite{Mastache:2016ahe} reports percentage errors of $h(n)$ of as much as 
$2\%$ for flat potentials, and up to $9\%$ when features are present.

It is significant that our Green's function $\mathcal{G}(n-m)$ depends 
only on the difference of its arguments, and we just need it over a range
of about ten e-foldings. Further, its Laplace transform is defined by 
relations (\ref{Laplacegreen}-\ref{INT}). Figure~\ref{Laplace} shows that 
there is only a single, simple pole on the real axis, somewhat below 
$s = 3$. If nothing else worked we could therefore evaluate $\mathcal{I}(
s_0 + i \omega)$ numerically for some $s_0 > 3$ and then numerically 
compute the inverse Laplace transform,
\begin{equation}
\mathcal{G}(n-m) = \frac1{2\pi i} \! \int_{-\infty}^{\infty} \!\!\!\! d\omega
\, e^{(s_0 + i\omega) n} \widehat{\mathcal{G}}(s_0 \!+\! i\omega;m) \; .
\end{equation}
No matter how time-consuming the computation proved, it would only need to be 
done once.

\section{Epilogue}

As its title suggests, this paper gives final expression to our formalism
for finding the tree order power spectra by evolving the norm-squared mode 
functions \cite{Romania:2012tb}. Considered purely as a numerical technique
this is more efficient than evolving the mode functions because it avoids 
keeping track of the rapidly fluctuating phase, and because it converges about
twice as fast. Nor is anything lost because the phase can be recovered through
expressions (\ref{ufromM}) and (\ref{vfromN}). Our formalism applies not only
to single-scalar inflation but also to any conformally related model, such as 
$f(R)$ inflation \cite{Brooker:2016oqa}, whose power spectra are numerically 
identical.

Section 2 reviews our formalism, and explains how to factor out arbitrary
approximate solutions (\ref{heqn}) and (\ref{geqn}). Section 3 then specializes 
to what we believe are the best choices (\ref{M0actual}-\ref{N0actual}) for these
approximate solutions. Our results (\ref{fullDh}-\ref{fullDR}) for the power
spectra are exact at this stage, with the nonlocal correction exponents 
$\tau[\epsilon](k)$ and $\sigma[\epsilon](k)$ given by (\ref{fullsigmatau}).

Section 5 makes the approximation that $\epsilon(n)$ is small, and that 
nonlinear effects can be dropped in the equations (\ref{heqn}) and (\ref{geqn}) 
for the residuals. This results in wonderfully simple, analytic approximations 
(\ref{tensornonlocal}-\ref{scalarnonlocal}) for how the nonlocal correction 
exponents depend upon $\epsilon(n)$. Figures~\ref{approximation} and 
\ref{StepTensor} exhibit the accuracy of these formulae, even for the model of
Figure~\ref{stepgeometry} which has prominent features. Figure~\ref{approximation} 
also demonstrates that the local slow roll approximation --- 
$\Delta^2_{\mathcal{R}}(k) \approx \frac{G H_k^2}{\pi \epsilon_k} \times 
C(\epsilon_k)$ --- breaks down badly when features are present, and that it 
systematically underestimates $\Delta^2_{\mathcal{R}}(k)$ even for models 
without features. The unmistakable conclusions are:
\begin{enumerate}
\item{That quantitative accuracy requires the nonlocal correction exponents 
$\tau[\epsilon](k)$ and $\sigma[\epsilon](k)$; and}
\item{That our approximations (\ref{tensornonlocal}-\ref{scalarnonlocal}) are
valid for any model which is consistent with the bounds on $r$ and on the
limits of possible features.}
\end{enumerate}

Section 5 explains how our approximation (\ref{scalarnonlocal}) can be used 
to reconstruct the geometry from the power spectra. (The scalar and its 
potential can be recovered from the formulae of footnote 1.) Further improvements
are needed for accurate reconstructions for derivatives of the first slow roll 
parameter, but the undifferentiated parameter is accurate to $\pm2.2\%$ and our 
errors for the Hubble parameter never exceed $0.04\%$. This seems much better 
than other techniques \cite{Kadota:2005hv,Barrow:2016wiy,Mastache:2016ahe}.

Our formalism has many applications because it gives explicit, analytic and
accurate approximations for how the power spectra depend functionally on the
geometry of inflation. For example, our expressions (\ref{fullDh}-\ref{fullDR})
imply an exact relation for the tensor-to-scalar ratio,
\begin{equation}
r(k) = 16 \epsilon_k \exp\Bigl[-\sigma[\epsilon](k) + \tau[\epsilon](k)
\Bigr] \; ,
\end{equation}
with no local, slow roll corrections. It should be an excellent approximation 
to drop $\tau[\epsilon](k)$ and employ the analytic approximation 
(\ref{scalarnonlocal}) for $\sigma[\epsilon](k)$.

We have already mentioned the necessity of including the nonlocal correction
exponent $\sigma[\epsilon](k)$ to correctly describe features. Our analytic
approximation (\ref{scalarnonlocal}) facilitates precision studies, limited
by the accuracy of the data rather than by the cumbersome connection to
theory. For example, the model of Figure~\ref{stepgeometry} was proposed \cite{Adams:2001vc,Mortonson:2009qv} to account for the deficit in the scalar 
power spectrum at $\ell \approx 22$, and the excess at $\ell \approx 40$,
which are visible in the data reported from both WMAP \cite{Covi:2006ci,
Hamann:2007pa} and PLANCK \cite{Hazra:2014goa,Hazra:2016fkm}. From 
Figure~\ref{approximation} we see that the resulting power spectrum indeed
has a deficit at $n \approx 172.3$, followed by an excess at $n \approx 172.8$.
However, there are weaker features at $n \approx 173.2$ and $n \approx 173.5$.
Do the data show any evidence for these weaker features? If not, to what degree
does their absence rule out the model of Figure~\ref{stepgeometry}? And what
sort of model do the data actually support? 
 
A particularly exciting application of our formalism is to exploit the
control it gives over how the mode functions depend upon $\epsilon(n)$ to
design a new statistic to cross-correlate features in the power spectrum
with non-Gaussianity. This has already been proposed in the context of models
with variable speed of sound \cite{Achucarro:2012fd,Torrado:2016sls} but it can 
now be done with precision for simple scalar potential models. Of course the 
idea is that non-Gaussianity measures self-interaction, which is what a step
in the potential provides. There may be an observable effect which is not 
resolvable by generic statistics but could be detected by a precision search.

Another application concerns the far future, after the tensor power spectrum
has been well resolved. Our analytic approximations 
(\ref{tensornonlocal}-\ref{scalarnonlocal}) quantify how the same derivatives of 
the first slow roll parameter lead to deviations from the local slow roll 
predictions for the tensor and scalar power spectra. Figure~\ref{approximation} 
shows that these deviations are strongly present in $\Delta^2_{\mathcal{R}}(k)$ 
for models with features. The associated tensor features are much weaker, but 
they can just be made out in Figure~\ref{StepTensor}. Demonstrating this 
correlation in the data would represent an impressive check on single-scalar 
inflation.

In the even farther future it may be possible to resolve one loop corrections
\cite{Woodard:2014jba}. Comparing these with theory obviously requires a precision
determination of the tree order effect, which is of course possible once the model
of inflation has been fixed. However, one also needs to be able to extract the
potentially large factors of $1/\epsilon(n)$ from the $\zeta$ propagator, and our
formalism is ideal for that.

\vspace{.5cm}

\centerline{\bf Acknowledgements}

This work was partially supported by the European Union's Seventh 
Framework Programme (FP7-REGPOT-2012-2013-1) under grant agreement 
number 316165; by the European Union's Horizon 2020 Programme
under grant agreement 669288-SM-GRAV-ERC-2014-ADG;
by NSF grant PHY-1506513; and by the UF's Institute for 
Fundamental Theory.

\end{document}